\documentclass{aa} 

\usepackage{graphicx}
\usepackage{txfonts}
\usepackage{soul,xcolor}
\setstcolor{red}
\usepackage{array} 
\usepackage{colortbl}
\usepackage{comment}
\usepackage{ulem}
\usepackage{hyperref} 
\usepackage{pdflscape}
\usepackage{multirow}
\newcommand{\rahms}[4]{$#1^{\rm h}#2^{\rm m}#3\mbox{$^{\rm s}\mskip-7.6mu.\,$}#4$} 
\newcommand{\decdms}[4]{$#1^{\circ}#2'#3\mbox{$''\mskip-7.6mu.\,$}#4$} 

\usepackage{booktabs}
\usepackage[switch]{lineno}

\begin{document}

   \title{Very long baseline interferometry detection of nearby ($<100$~pc) young stars    }

   \subtitle{Pilot observations}

   \author{Sergio A. Dzib
          \inst{1}
          \and
          Laurent Loinard\inst{2}
          \and
          Ralf Launhardt\inst{3}
          \and
          Jazm\'{i}n Ord\'o\~nez-Toro\inst{2}
          }

   \institute{Max-Planck-Institut f\"ur Radioastronomie, Auf dem H\"ugel 69, D-53121 Bonn, Germany
   \and
    Instituto de Radioastronom\'{\i}a y Astrof\'{\i}sica, Universidad Nacional Aut\'onoma de M\'exico, 58090, Morelia, Michoac\'an, Mexico
    \and 
    Max-Planck-Institut f\"ur Astronomie, K\"onigstuhl 17, D-69117 Heidelberg, Germany}


 
  \abstract
  {To increase the number of sources with very long baseline interferometry (VLBI) astrometry available for comparison with the {\it Gaia} results,
  we have observed 31 young stars with recently reported radio emission.
  These stars are all in the {\it Gaia} DR3 catalog and were suggested, on the basis of conventional interferometry observations, to be nonthermal radio emitters and are therefore good candidates for VLBI detections. The observations were carried out with the Very Long Baseline Array (VLBA) at two epochs separated by a few days and yielded ten detections (a roughly $30$\% detection rate).
  Using the astrometric {\it Gaia} results, 
  we extrapolated the target positions to the epochs of our radio observations
  and compared them with the position of the radio sources.
  For seven objects, the optical and radio positions are coincident within five times their combined position errors. Three targets, however, have position discrepancies 
  above eight times the position errors, indicating different emitting sources at optical and radio wavelengths. In one case, the VLBA emission is very likely associated with a known companion of the primary target. In the two other cases, we associated the VLBA emission with previously unknown companions, but further observations will be needed to confirm this.
  }


   \maketitle
%

\section{Introduction}

Stellar radio astrometry with very long baseline interferometry (VLBI) {at centimeter wavelengths}
achieves angular resolutions of a few milli-arcseconds and an astrometric accuracy on the order of $0.1$~mas 
\citep[e.g.,][]{reid2014}. 
For young stars, VLBI radio astrometry has primarily been used 
to better constrain star-forming region distances, which were poorly known at the time 
or were debated due to discrepant results from indirect methods 
\citep[e.g.,][]{loinard2005,loinard2007,dzib2010,ortiz2017a}.  
In young stars, radio emission detectable with VLBI is
nonthermal and is produced in stellar coronae \citep{gudel2002} that typically extend 
up to a few radii and remain unresolved with 
VLBI observations at centimeter wavelengths. Thus, the astrometry 
derived from these radio observations directly traces the stellar motion.
Recently, the {\it Gaia} mission \citep{gaia2016,gaiadr32021,gaiadr3_2023} has been delivering astrometry 
of over a billion stars, including many nearby young stars. Not surprisingly, {\it Gaia} 
results have confirmed the astrometric results previously reported using 
the VLBI technique \citep[e.g.,][]{ortiz2018,galli2019}.

Gaia and VLBI astrometry accuracies are comparable, so the two techniques are complementary
\citep[e.g.][]{dzib2021}. Comparing trigonometric parallaxes measured
with both can constrain the {\it Gaia} zero-point offset 
\citep{lindegren2018,lindegren2020,riess2018,zinn2019,xu2019}.
{Also, high precision VLBI astrometry allows for the linking of the optical stellar reference frames to the radio extragalactic reference frame since VLBI stellar astrometry is relative to them \citep{lestrade1999}.} 
Recently, \citet{lindegren2020} analyzed 41 stars in the {\it Gaia} 
catalog with existing VLBI stellar astrometry. He noticed 
discrepancies between the VLBI and {\it Gaia}-derived astrometry and rejected 15 stars 
for further 
analysis. These rejected stars are binary and multiple 
systems that do not comply with 
the linear motion prescription currently used for {\it Gaia} astrometry. 
From the remaining 26 stars, he estimated 
the rotation of the \textit{Gaia} bright  
stars ($G\leq13$~mag) reference frame with respect to the distant quasar 
frame (the so-called \textit{Gaia} Celestial Reference Frame). \citet{lindegren2020} 
recognized that stellar VLBI astrometry can provide independent
astrometric results that could be used to test results from current and 
future {\it Gaia} releases. He also noted that for this purpose, it will be necessary 
to re-observe stars with VLBI and identify new stars that can be detected with the VLBI technique.
There are, however, few stars with suspected nonthermal radio emission
that have not already been observed with the VLBI technique.

Looking to increase the number of stars with both VLBI and {\it Gaia} astrometry, 
the main goal of the present work is to obtain VLBI observations of 
nearby young stars with suspected nonthermal radio emission and compare their 
positions with those reported by {\it Gaia}. 
Here we report VLBI observations of 31 nearby ($<100$~pc) young stars ($<100$~Myr)
that were recently reported to be radio emitters by \citet{launhardt2022}.
These authors also suggested that the radio emission is likely of nonthermal
nature and could thus be detected with VLBI observations.
The stars have optical counterparts, and most of them are bright stars
in the {\it Gaia} catalog. In Table~\ref{tab:tar}, these stars are listed
together with their astrometric solution from {\it Gaia} DR3  \citep{gaiadr32021,gaiadr3_2023} 
and other known properties.


\begin{landscape}
\setlength{\tabcolsep}{5.7pt}
\begin{table}[p]
\centering
\scriptsize
\caption{Target sources.}
\renewcommand{\arraystretch}{1.3}
\begin{minipage}{\linewidth}
\begin{tabular}{lcccccccccccc|cc}\hline\hline
{Star ID} & {R.A. (J2016.0)} & {Dec. (J2016.0)} &$\pi$ & {Dist.}  & $\mu_\alpha\cdot\cos{(\rm \delta)}$& $\mu_\delta$&{spT} & {age}& {age}&G&Dates for Max. & Obs. & Phase  & Separation\\
{} & {($^\circ$)} & {($^\circ$)} & {(mas)} &{(pc)}  & (mas~yr$^{-1}$) & (mas~yr$^{-1}$) & {} &{(Myr)}& {ref.} & (mag)  &Parallax& Group& Cal. & (${^\circ}$)\\
\hline
BD+17\,232  &$024.4144646604$&$+018.5922019429$&$18.95\pm00.03$&$52.77\pm00.09$&$73.79\pm00.04$&$-43.35\pm00.03$.      & K3VE & 10 & 1&10.18 &  Jan 16 - Jul 18 & A & J0139+1753 & 0.9\\
HIP\,12545 &$040.3582211403$&$+005.9881913726$&$22.58\pm00.02$&$44.30\pm00.04$&$79.07\pm00.02$&$-56.70\pm00.02$.      & K6Ve &  10& 2& 9.82 &Feb 2 - Aug 4 & B& J0239+0416 & 1.8 \\
HIP\,12635 &$040.5877605186$&$+038.6220662083$&$20.09\pm00.02$&$49.78\pm00.05$&$82.11\pm00.02$&$-107.55\pm00.02$.     & K2V  & 149& 3 & 9.92 &Feb 2 - Aug  5 & B& J0253+3835&2.1\\
\rowcolor{lightgray}
\emph{V\,875\,Per}  &$043.0736090107$&$+036.2798896433$&$04.07\pm00.02$&$245.92\pm01.08$&$52.87\pm00.02$&$-38.55\pm00.02$& K2IV  &  63&4 &10.45 & Feb 4 - Aug  7& B & J0246+3536 & 1.4 \\
{ TYC\,3301-2585-1}\tablefootmark{a} &$043.9317736370$&$+047.7795639137$&$19.97\pm00.02$&$50.07\pm00.05$&$85.50\pm00.02$&$-79.53\pm00.02$& K5Ve & 42 &3 &10.71 &Feb 5 - Aug  8& B & J0303+4716 & 1.4\\
\rowcolor{lightgray}
\emph{ HD\,22213} &$053.5684734519$&$-012.0688463141$&$19.46\pm00.02$&$51.38\pm00.05$&$74.46\pm00.02$&$-34.36\pm00.01$& G7V 	&  45 &3 &8.58 &Feb 15 - Aug  18 & C & J0336-1302 & 1.1\\
{ HD\,23524}\tablefootmark{b} &$057.0963206572$&$+052.0375180643$&$19.35\pm01.62$&$51.85\pm04.50$&$61.87\pm01.98$&$-70.99\pm01.67$& K1V & 42&3 &8.62 &Feb 18 - Aug  21 &{\bf D} & J0350+5138 &0.5 \\
HD\,24681  	&$058.8352088930$&$-001.7296295356$&$17.94\pm00.02$&$55.73\pm00.06$&$43.50\pm00.02$&$-91.47\pm00.02$& G8V & 149 &3 &8.79 & Feb 20 - Aug  23&{\bf C} &J0351-0301 & 1.7 \\
\rowcolor{lightgray}
\emph{ HD\,285281}  &$060.1294810416$&$+019.5890872952$&$07.34\pm00.02$&$136.22\pm00.43$&$04.04\pm00.03$&$-12.36\pm00.02$& K1  & 1.5 &4 &9.97 &Feb 21 - Aug  24&D& J0401+2110&1.6\\
HD\,284135\tablefootmark{c} 	&$061.4191041862$&$+022.8032778542$&$12.50\pm01.00$&$80.69\pm06.55$&$05.50\pm01.00$&$-14.70\pm01.00$&G3V: &  1.5 &1 &9.17 &Feb 22 - Aug  25&D& J0412+2305 & 1.7\\
{ HD\,31281}  	&$073.7900713168$&$+018.4419023094$&$08.25\pm00.03$&$121.29\pm00.43$&$-2.51\pm00.04$&$-16.10\pm00.02$& G1V: &  1.5 &5 &9.06  &Mar 6 - Sep 6&E& J0449+1754 & 1.5\\
BD-08\,995 	&$074.7023574082$&$-008.7277965955$&$11.47\pm00.02$&$87.15\pm00.12$&$25.83\pm00.01$&$-21.61\pm00.01$& K0V	&  42 &3,6 &10.00 &Mar 7 - Sep 7&E& J0457-0849 & 0.4\\
HD\,286264  &$075.2054343407$&$+015.4499325911$&$18.83\pm00.02$&$53.09\pm00.04$&$18.31\pm00.02$&$-58.72\pm00.01$& K2IV &24 &3 &10.37 &Mar 7 - Sep 8&E& J0459+1438 & 0.9\\
HD\,293857\tablefootmark{d}  &$077.7903401655$&$-004.1820005272$&$12.82\pm01.00$&$78.45\pm06.24$&$18.70\pm01.00$&$-57.50\pm01.00$& G8V &  24  &3 &9.04 &Mar 10 - Sep 10&E&J0505-0419&1.3 \\
TYC\,713-661-1 &$084.2085922983$&$+013.6317744334$&$17.61\pm00.02$&$56.77\pm00.06$&$04.90\pm00.02$&$-109.84\pm00.01$& K0V& 149 &3 &10.16 &Mar 16 - Sep 16&F& J0539+1433 &1.2\\
TYC\,5925-1547-1 &$084.8465611076$&$-019.5583471893$&$14.30\pm00.17$&$69.95\pm00.82$&$21.30\pm00.13$&$-48.63\pm00.16$& K1V & 150 & 3,6&10.28  &Mar 16 - Sep 17&F& J0536-2005 &0.9\\
\rowcolor{lightgray}
\emph{ AI\,Lep}    	&$085.0864740191$&$-019.6697559694$&$13.86\pm00.01$&$72.16\pm00.07$&$19.32\pm00.01$&$-11.69\pm00.01$& G2V & 42 &3 &8.94 &Mar 17 - Sep 17&F& J0536-2005 &1.0\\
\rowcolor{lightgray}
\emph{ HD\,62237}       &$115.6106888262$&$-016.2834934596$&$08.02\pm00.10$&$124.76\pm01.60$&$-9.44\pm00.09$&$-13.21\pm00.09$& G5V 	& 42 &3,6 &9.39 &Apr 14 - Oct 16 &G& J0746-1555 & 1.0\\
\rowcolor{lightgray}
\emph{ SAO\,135659}&$123.4623602751$&$-007.6403837042$&$19.97\pm00.20$&$50.07\pm00.51$&$-32.15\pm00.23$&$-32.31\pm00.19$& K0& 42 &3 &9.01&Apr 22 - Oct 23&G&J0808-0751 & 1.4\\
WDS\,09035+3750\,B	 &$135.8623936283$&$+037.8407266874$&$32.75\pm00.12$&$30.54\pm00.11$&$-65.85\pm00.11$&$-163.11\pm00.08$& M3-6V & 42  &3 &6.91 &May 4 - Nov 5&H& J0850+3747 &2.6 \\
\rowcolor{lightgray}
\emph{ HD\,82159}	     	&$142.6484056528$&$+010.6016777097$&$28.23\pm00.04$&$35.43\pm00.05$&$-200.06\pm00.04$&$-12.88\pm00.03$& G9V   	&   150 &7 &8.41 &May 11 - Nov 11&H&J0935+0915&1.8 \\
\rowcolor{lightgray}
\emph{ HD\,82558}     		&$143.1054112092$&$-011.1844843913$&$54.74\pm00.02$&$18.27\pm00.01$&$-248.04\pm00.02$&$34.28\pm00.02$& K0V	&  43 &8 &7.50 &May 12 - Nov 12&H& J0933-1139 &0.6 \\
GJ\,2079     	 	&$153.5792772472$&$+021.0741011408$&$42.74\pm01.00$&$23.41\pm00.55$&$-119.55\pm01.00$&$-191.19\pm01.00$& M0.7V	&  24 &3 &9.36 &May 23 - Nov 23&H& J1016+2037 &0.7\\
HD\,135363  		&$226.9819941355$&$+076.2014908036$&$33.71\pm00.05$&$29.66\pm00.05$&$-127.19\pm00.08$&$165.47\pm00.07$& G5V   	&   45 &9 &8.39 &Feb 8 - Aug  11&I& J1448+7601&1.2\\
UCAC4\,832-014013  &$226.9860077519$&$+076.2338226259$&$33.71\pm00.03$&$29.66\pm00.02$&$-127.64\pm00.04$&$163.10\pm00.04$& M4.5V   &  ...  &... &12.40&Feb 8 - Aug  11&I&J1448+7601&1.2 \\
\rowcolor{lightgray}
\emph{ HD\,199143}	&$313.9489042861$&$-017.1144445844$&$21.78\pm00.03$&$45.91\pm00.07$&$55.73\pm00.03$&$-58.67\pm00.03$& F8 & 24 &3 &7.32 &May 2 - Nov 3&J& J2107-1708 & 2.7 \\
HD\,358623	 	&$314.0116622465$&$-017.1818893172$&$21.70\pm00.02$&$46.08\pm00.05$&$57.32\pm00.02$&$-62.20\pm00.02$& K6Ve  	&   24 &3 &9.90{*}&May 2 - Nov 3&J& J2107-1708 & 2.7\\
SAO\,50350		&$315.1960688588$&$+045.5029486058$&$19.31\pm00.01$&$51.78\pm00.03$&$38.84\pm00.01$&$08.50\pm00.01$& F8 	&   200 &8 &8.36 &May 4 - Nov 4&J& J2102+4702 & 1.6\\
\rowcolor{lightgray}
\emph{ GJ\,4199}     	 	&$322.7577934607$&$+023.3347377313$&$41.29\pm00.02$&$24.22\pm00.01$&$134.65\pm00.01$&$-144.89\pm00.01$& K5Ve 	&   149 &3 &8.85 &May 11 - Nov 12&K& J2125+2442 & 1.9\\
SAO\,51891 		&$335.0299106318$&$+049.5033014554$&$29.07\pm00.02$&$34.40\pm00.02$&$93.24\pm00.02$&$07.37\pm00.02$& K1V   	&   37 &8 &8.22 & May 24 - Nov 24&K&J2201+5048&3.2 \\
SAO\,108142 	&$341.1734897431$&$+017.9047178334$&$20.85\pm00.07$&$47.97\pm00.17$&$82.49\pm00.08$&$-82.11\pm00.07$& K0    	&   149 &3 &9.17 &May 31 - Dec 1&K&J2242+1741&0.5\\
\hline\hline
\label{tab:tar}
\end{tabular}
\tablefoot{Stars with a detected radio source in this work are highlighted. The positions, parallax, and proper motion were obtained from the {\it Gaia} DR3 catalog \citep{gaiadr32021,gaiadr3_2023}, with some exceptions. \\
\tablefoottext{a}{Hereafter TYC\,3301. }
\tablefoottext{b}{Star in {\it Gaia} DR3 without parallax and proper motions. Position is from {\it Gaia} DR3, and parallax and proper motions are taken from the Hipparcos catalog \citep{leeuwen2007}.}
\tablefoottext{c}{Star in {\it Gaia} DR3 catalog without parallax and proper motions. Position is from {\it Gaia} DR3, proper motions are from Tycho-2 catalog \citep{hog2000} and distance information from \citet{carpenter2009}.}
\tablefoottext{d}{Star in {\it Gaia} DR3 catalog without parallax and proper motions. Position is from {\it Gaia} DR3, proper motions are from the Tycho-2 catalog, and distance information is from \citet{dasilva2009}. The last column lists the angular separation between the target source and the phase calibrator.}
}
\vspace{0.2cm}

{\raggedright \footnotesize
      Age references: (1) \citet{galicher2016}, (2) \citet{weise2010}, (3) \citet{bell2015}, (4) \citet{carpenter2009}, (5) \citet{KH1995}, (6) \citet{dasilva2009}, (7)  \citet{desidera2015}, (8)  \citet{stanford2020}, and (9) \citet{montes2001}. \par}
\end{minipage}
\end{table}
\end{landscape}
%

\section{Observations}

The observations were obtained {at 20 epochs in the first half of 2021} with the Very Long Baseline Array (VLBA) using the
C-band receivers ($\lambda=6.0$~cm; $\nu=4.8$~GHz). This band was selected 
as a compromise between the angular resolution, strength of radio fluxes, 
and the high sensitivity offered by this band. We used the newly offered 
recording bit rate of 4~Gbps, which improves the sensitivity 
of the array by a factor of $\sqrt{2}$ with respect to the previous system
for an equal observing time. {For the present observations, about 45 minutes were spent on each target source per observation, expecting a noise level of 19~$\mu$Jy~beam$^{-1}$, while with the previous system was 27~$\mu$Jy~beam$^{-1}$.}

\subsection{Observation strategy}

The radio emission from the stars in our sample is known to be rapidly variable, and thus we
scheduled two observing runs separated by a few days to increase the chances
of detection. The observations were scheduled close to the maximum trigonometric parallax 
extension in right ascension so that they could be combined with future observations to measure the 
trigonometric parallaxes of the detected sources. The observing dates are listed individually for all 31 
radio stars in Table~\ref{tab:epochs}. 

As the target sources are weak, phase-referencing was used
\citep[e.g., ][]{lestrade1990,beasley1995}. This means that the target observing 
scans (2~min.) are bracketed with scans (1~min.) on a relatively bright quasar with an 
angular separation of $\lesssim 3^{\circ}$. The phase corrections determined for the quasar 
were transferred to the target so the target source positions are referenced to the quasars. 
The use of this technique, including antenna slewing time, increases the total telescope 
time by 60\% (for instance, for 30 minutes on source, we would need 48 minutes in total). Additionally, 
to reduce the impact of errors in the troposphere models used by the correlator, we
observed so-called geodetic blocks 
(i.e.,\ $\sim15$-minute 
blocks at the beginning and/or middle and/or end of the observation) dedicated 
to observations of five to six bright quasars distributed over the entire sky {\citep{reid2004}}.

To reduce the overheads caused by the geodetic blocks, we organized the targets
into 11 groups of sources with similar right ascensions (and therefore also similar dates 
of maximum parallax elongation). We named these groups using alphabetic letters from A to K. We have one group of one target, three groups of two targets, 
four groups of three targets, and three groups of four targets (see Table~\ref{tab:epochs}). For the groups
with one and two targets, we used two geodetic blocks (at the beginning and the end of the session). 
For groups with three or four targets, we used three geodetic blocks (at the beginning, middle, 
and end of the sessions). A second advantage of the grouping strategy is that we could schedule 
the observations such that the $(u,\,v)$ plane coverage is maximized. This was achieved by intertwining 
several blocks of 24 minutes of the different sources with similar right ascension throughout  
the duration of the session.

\begin{table}
\small
\begin{center}
\renewcommand{\arraystretch}{1.0}
\caption{Observed epochs.}
\begin{tabular}{c|cccccc}\hline\hline
&    Epoch 1 &  Epoch 2  & No. observed\\
Group& yyyy/mm/dd &  yyyy/mm/dd& targets\\ 
\hline
A & 2021/07/15  & 2021/07/17 & 1\\
B & 2021/01/22  & 2021/01/24 & 4\\
C & 2021/02/10  & 2021/02/13 & 2\\
D& 2021/02/12   & 2021/02/14 & 3\\
E & 2021/03/09  & 2021/03/10 & 4\\
F & 2021/03/12  & 2021/03/13 & 3\\
G & 2021/04/14  & 2021/04/16 & 2\\
H & 2021/05/16  & 2021/05/23 & 4\\
I & 2021/01/29  & 2021/01/30 & 2\\
J & 2021/05/07  & 2021/05/10 & 3\\
K & 2021/05/21  & 2021/05/22 & 3\\
\hline\hline
\end{tabular}\label{tab:epochs}
\end{center}
\tablefoot{Columns are (from left to right) the name of the group, the two epochs of the observation in civil dates, and the number of targets included in the corresponding group. }
\end{table}

\subsection{Calibration and imaging}

The data were edited and calibrated using the software Astronomical 
Image Processing System \citep[AIPS;][]{greisen2003}. After a detailed
inspection, flawed data were removed. The data were calibrated following the 
standard schemes plus a calibration of the group delay {(i.e., the rate of phase change with frequency). The calibration
of the group delay was used to further reduce systematic errors caused by tropospheric 
zenith delays and ionospheric content delays \citep{reid2004}. With this purpose, we additionally observed ICRF quasars distributed over the sky over a wide range of frequencies \citep[also known as the geodetic blocks; ][]{reid2004}. First, the geodetic block data were calibrated using a standard procedure. Second, the clock delays and tropospheric terms were derived using the AIPS task DELZN and applied to the target data set using the CALIB.}
A detailed description of the calibration strategy has been summarized by 
\citet{loinard2007,dzib2010} and \citet{ortiz2017a}.

\begin{figure*}[t!]
\centering
\includegraphics[width=1.0\textwidth, trim= 45 20 20 0,clip]{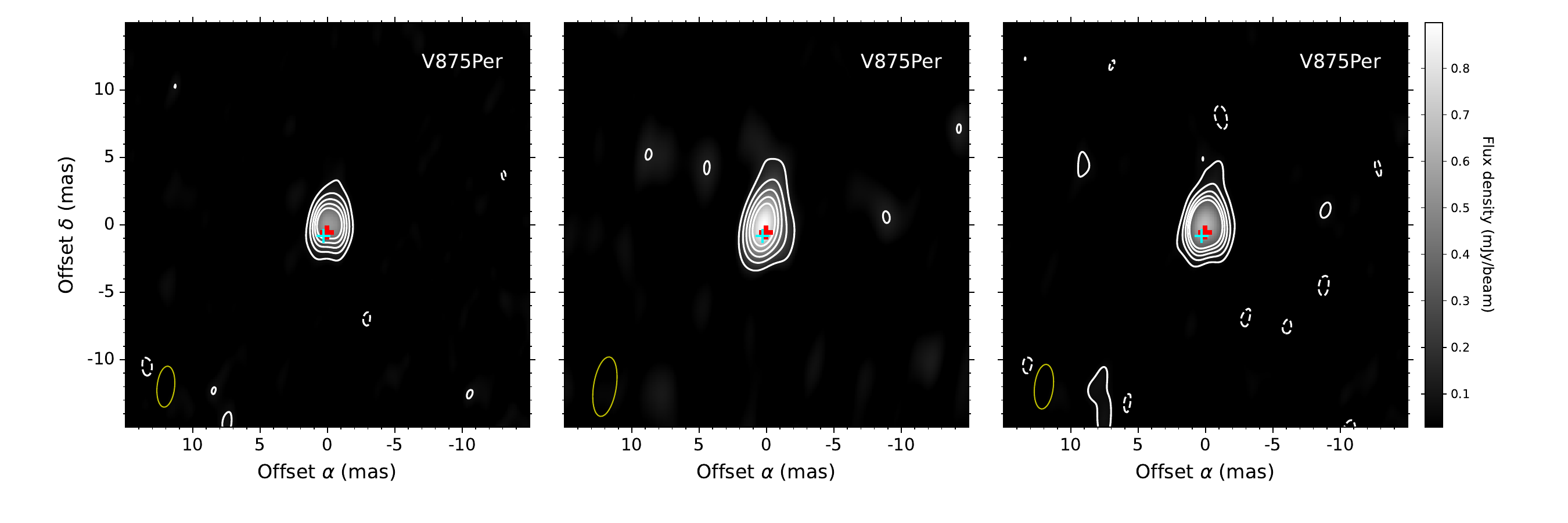}
\caption{VLBA images of the radio source related to V~875~Per. The images are centered in the position of the radio source as detected in the first epoch (see Table~\ref{tab:am}). Images, from
left to right, correspond to epochs 1, 2, and the combination of both epochs. Contour levels are --3, 3, 6, 9, 12, and 15 times the noise level of the image as listed in Table~\ref{tab:Imr}. The 
yellow open ellipse in the bottom-left corner represents the size of the synthesized beam of the image as listed in Table~\ref{tab:Imr}. The predicted optical position in epochs 1 and 2 are shown as red and cyan crosses, respectively.}\label{fig:NF}
\end{figure*}

After calibration, the data were imaged in two steps. First, 
for the detection experiment, the $(u,\,v)$ data of the two epochs for 
each target were combined using the task DBCON in AIPS. 
This was done with the purpose of 
decreasing the noise level in the images (by a factor of $\sqrt{2}$) 
and increasing the chances of detections. 
Some of the target 
sources have very high proper motion (up to $250$~mas~yr$^{-1}$, 
see Table~\ref{tab:tar}), so before concatenating the data of the two observed epochs, 
we corrected the position of the second epoch to coincide with that of the first epoch. 
This procedure is described in Appendix~\ref{app:1}.
After the concatenation of the two epochs, we imaged large areas around the expected positions of the target 
sources. These images are $2''\times2''$ with 4096 pixels, 500$\,\mu$as each, along both directions, and they 
used a natural weighting scheme 
({\it Robust=5} in AIPS).\footnote{The robust scheme weights the 
visibilities considering the length of the baselines. A natural 
weighting scheme can be set with {\it Robust=5} in AIPS and gives more 
weight to the interferometer's shortest baselines. A uniform 
scheme weights all the baselines equally and is set with 
{\it Robust=--5} in AIPS. The first scheme resulted in images with better 
sensitivity, and the second scheme resulted in the smallest 
synthesized beam but with higher noise.}

In the second step, a new image was obtained for each source around the peak emission 
on the first image. These new images are  $0.1''\times0.1''$ with 1024 pixels, 100$\,\mu$as each, 
and they also used natural weighting. We consider a source detection when the peak value
has a signal-to-noise ratio (SNR) $>6.5$ in this last image
\citep[see discussion by][]{forbrich2021}. 
The final images were produced at the position of the detected sources in each 
of the data sets of individual epochs using the parameters of the previous 
image. Only in two cases was the detected emission strong enough (with  
S/N$>20$) that we could produce final images using an intermediate weighting 
scheme between natural and uniform 
\citep[{\it Robust=0} in AIPS;][]{briggs1995} and obtain a better 
position measurement. The rms noise levels in the final images are
$\sim20$~$\,\mu$Jy beam$^{-1}$, similar to the VLA images reported
by \citet{launhardt2022}. 
The flux densities and positions of the detected sources were measured using a 
two-dimensional Gaussian fitting procedure (task JMFIT in AIPS).
Statistical errors resulting from JMFIT are given for each detection in columns (4) and (6) in Table~\ref{tab:am}. 
Additionally, systematic residual phase errors caused by the transfer of the 
calibration from the phase calibrator to the target were expected. 
These errors are on the order of $0.1$~mas per degree
of the angular separation \citep[e.g.,][]{reid2017}. Given that 
the maximum angular separation between our targets and their phase 
calibrators is 3$\rlap{.}^\circ2$, our systematic errors are expected to be
$\leq0.32\,$mas. These systematic errors were calculated specifically for each detected target 
source. {Overall, we note that the position errors for weak sources ($S_{\nu,{\rm peak}}<200\,\mu$Jy beam$^{-1}$) are dominated by the statistical errors, and for the strongest sources, they are dominated by the systematic residual phase errors. }

\section{Results}

Radio emission was detected for ten of our 31 targets. The properties of the 
final images and the flux densities are listed in Table~\ref{tab:Imr}, and the positions 
of the radio sources are given in Table~\ref{tab:am}. 

In the plots of Figure~\ref{fig:NF}, we show an example of a detected radio source,
and the remaining images of other detected radio sources are shown in 
Figure~\ref{fig:NF2}.
In most cases, single sources are detected in the images.
A possible exception is source HD~199143, where two emission peaks separated by a few 
milli-arcseconds are detected in the first epoch. Given the low signal-to-noise ratio 
of the second peak, however, it is not possible to 
clearly establish that the second peak is a real source.
Sources HD~82558 and HD~82159 are 
single sources in both epochs, but their positions change appreciably from one epoch
to the next due to their high proper motion. As mentioned earlier, the combined image
aligns both epochs to the position in the first image.
The radio sources are rapidly variable, and in several 
cases, their flux density changes by a factor larger than two in a few days. This high variability 
was also noticed in the VLA observation \citep{launhardt2022}
and is typical of stellar sources with magnetic activity \citep[e.g.,][]{gudel2002}.

\begin{table}[!h]
\scriptsize
\caption{Image properties and measured fluxes of detected radio sources.}
\renewcommand{\arraystretch}{1.25}
\setlength{\tabcolsep}{4.5pt}
\begin{tabular}{r|ccccccccc}\hline\hline
          &          &       & Synthesized beam & &\\
          &          & Noise & $\theta_{\rm maj}\times\theta_{\rm min}; {\rm P.A.}$ & $S_{\nu,{\rm Int.}}$&$S_{\nu,{\rm Peak.}}$\\
{Star ID} & Epoch & ($\mu$Jy bm$^{-1}$) & (mas $\times$ mas; $^\circ$) & ($\mu$Jy) & ($\mu$Jy bm$^{-1}$) \\
\hline
V\,875\,Per &B1&26& 3.1$\times$1.3; $-5.9$ & $1042\pm66$ & $585\pm26$\\
& B2 & 42 & 4.5$\times$1.7; $-9.0$ & $1225\pm86$ & $885\pm42$ \\
& B1+B2 & 23 & 3.3$\times$1.4; $-6.5$ & $1156\pm56$ & $667\pm23$\\ \cline{2-6}
HD\,22213 & C1 & 24 & 3.6$\times$1.3; $+1.5$ & $338\pm88$ & $142\pm24$ \\
& C2 & 26 & 3.5$\times$1.1; $+9.7$ & $710\pm54$ & $552\pm26$ \\
& C1+C2 &  17 & $3.9\times1.4$; $+9.9$ & $450\pm40$ & $297\pm17$\\ \cline{2-6}
HD\,285281 & D1 & 17 & 4.0$\times$1.7; $-6.1$ & $270\pm62$ & $96\pm17$ \\
& D2 & 24 & 3.9$\times$1.6; $-3.6$ & $152\pm47$ & $141\pm26$ \\
& D1+D2 & 13 & 3.9$\times$1.6; $-4.9$ & $170\pm33$ & $103\pm13$\\ \cline{2-6}
AI\,Lep & F1 & 26 & 4.7$\times$1.8; $+6.7$ & $541\pm59$ & $387\pm27$ \\
& F2 & 15 & 5.0$\times$1.8; $+0.1$ & $105\pm33$ & $69\pm15$ \\
& F1+F2 & 12 & 4.8$\times$1.8; $+1.9$ & $220\pm36$ & $144\pm16$\\\cline{2-6}
HD\,62237 & G1 & 21 & 4.3$\times$1.6; $-1.8$ & $415\pm65$ & $190\pm21$ \\
& G2 & 19 & 4.3$\times$1.6; $-0.2$ & $172\pm35$ & $164\pm19$ \\
& G1+G2 & 15 & 4.3$\times$1.6; $-1.0$ & $280\pm38$ & $162\pm15$\\ \cline{2-6}
SAO\,135659 & G1 & 20 & 4.2$\times$1.6; $-2.7$ & $648\pm41$ & $540\pm21$ \\
& G2 & 22 & 4.2$\times$1.7; $-0.4$ & $524\pm41$ & $444\pm22$ \\
& G1+G2 & 16 & 4.2$\times$1.7; $-1.6$ & $586\pm31$ & $489\pm16$\\ \cline{2-6}
HD\,82159 & H1 & 66 & 3.9$\times$1.6; $+3.2$ & $980\pm138$ & $734\pm66$ \\
& H2 & 47 & 4.0$\times$1.7; $+1.5$ & $470\pm76$ & $503\pm47$ \\
& H1+H2 & 38 & 3.9$\times$1.6; $+2.2$ & $669\pm74$ & $577\pm39$ \\ \cline{2-6}
HD\,82558 & H1 & 38 & 4.0$\times$1.6; $+0.8$ & $462\pm77$ & $362\pm38$\\
& H2 & 25 & 4.2$\times$1.6; $-1.2$ & $219\pm40$ & $230\pm25$ \\
& H1+H2 & 22 & 4.1$\times$1.6; $-0.3$ & $365\pm65$ & $260\pm27$\\ \cline{2-6}
HD\,199143 & J1 & 30 & 5.1$\times$1.9; $-5.4$ & $280\pm81$ & $155\pm31$ \\
& J2 & 20 & 5.7$\times$1.6; $-9.7$ & $162\pm57$ & $81\pm20$ \\
& J1+J2 & 17 & 5.4$\times$1.7; $-8.2$ & $221\pm64$ & $113\pm17$\\ \cline{2-6}
GJ\,4199 & K1 & 19 & 4.0$\times$1.8; $-15.9$ & $350\pm47$ & $210\pm19$ \\
& K2 & 24 & 4.0$\times$1.8; $-12.0$ & $273\pm51$ & $192\pm24$ \\
& K1+K2 & 14 & 4.0$\times$1.8; $-13.8$ & $306\pm33$ & $204\pm15$\\ 
\hline\hline
\end{tabular}
\label{tab:Imr}
\tablefoot{Columns are (from left to right) the target star name, the epoch (for this table the epoch ID is constructed using the group name and the epoch number from Table 2), the image noise level, the synthesized beam size in units of milli-arcseconds, and the integrated and peak flux densities. 
}
\end{table}

        \begin{table*}[!h]
                \scriptsize
                \begin{center}
                \caption{Astrometry of detected radio sources.}
                \setlength{\tabcolsep}{2.7pt}
                \renewcommand{\arraystretch}{1.3}
                \begin{tabular}{lccccccccccccc}\hline\hline
                         & & & $\sigma_{\alpha}$ & & $\sigma_{\delta}$ &$\Delta\alpha^*_{\rm radio}$ & $\Delta\delta_{\rm radio}$ &$\Delta\theta_{\rm radio}$& $\Delta\alpha^*$ & $\Delta\delta$ & $\Delta\theta$& \normalfont{$\sigma_{\rm sys}$} & $\Delta\lambda$\\
                {Star ID}       & Epoch&{R.A.} &($\mu$s)  &{Dec} &($\mu$as)&($''$)&($''$)&($''$)&(mas)&(mas)&(mas)&(mas)&(au)\\
                (1)&(2)&(3)&(4)&(5)&(6)&(7)&(8)&(9)&(10)&(11)&(12)&(13)&(14)\\
\hline
V\,875\,Per &B1   &\rahms{02}{52}{17}{687958} & 3.3     &\decdms{+36}{ 16}{ 47}{407545}&    60 & $0.06\pm0.10$&$0.02\pm0.07$&$0.06\pm0.10$& $0.19\pm0.16$ & $-0.61\pm 0.12$ & $0.6\pm 0.1$ & 0.14& $ 0.10\pm  0.06$ \\
                        &B2   &\rahms{02}{52}{17}{687982}       &3.8&\decdms{+36}{ 16}{ 47}{407500}&       85 & & &\\
                        &B1+B2&\rahms{02}{52}{17}{687968} & 2.7 &\decdms{+36}{ 16}{ 47}{407501}& 49 & &\\
HD\,22213   &C1   &\rahms{03}{ 34}{ 16}{458216}&        22&\decdms{     -12}{ 04}{ 08}{026964}&       374& $-0.09\pm0.10$&$-0.07\pm0.14$&$0.11\pm0.12$&       $-1.0\pm0.3$ & $+0.2\pm 0.4$ & $1.1\pm0.3$ & 0.11&$ 0.05\pm 0.02$ \\
                        &C2   &\rahms{03}{34}{ 16}{458292} &    1.9&\decdms{    -12}{ 04}{ 08}{026581}&       79&&\\
                        &C1+C2&\rahms{03}{ 34}{ 16}{458287} & 2.5&\decdms{ -12}{ 04}{ 08}{026752} & 98 &&\\
HD\,285281&D1&     \rahms{04}{00}{31}{076436}&  14&\decdms{     +19}{ 35}{ 20}{650376} &   580& $-0.19\pm0.18$&$0.04\pm0.19$&$0.19\pm0.18$&        $+0.3\pm0.2$ & $+0.8\pm 0.6$ & $0.9\pm0.5$ & 0.16& $0.12\pm0.06$ \\
                        &D2   &\rahms{04}{00}{31}{076421}&      8.6&\decdms{+19}{ 35}{ 20}{648346} &      332&\\
                        &D1+D2&\rahms{04}{00}{31}{076421} & 7.2 &\decdms{ +19}{ 35}{ 20}{649016 }&317 &\\
AI\,Lep     &F1 &\rahms{05}{40}{20}{759902} &   4.8     &\decdms{-19}{ 40}{ 11}{183095}& 152& $0.55\pm0.37$&$0.32\pm0.42$&$0.64\pm0.39$&$+0.2\pm0.1$ & $+0.4\pm 0.2$ & $0.5\pm 0.1$ &0.10& $ 0.04\pm  0.01$ \\
                        &F2&\rahms{05}{ 40}{ 20}{759861}&       18&     \decdms{-19}{ 40}{ 11}{182392}&       465&    \\
                        &F1+F2&\rahms{05}{ 40}{ 20}{759892}&    8.6&\decdms{    -19}{ 40}{ 11}{182835}&       238&    \\
HD\,62237&G1    &\rahms{07}{ 42}{ 26}{562160}&  8.4&\decdms{-16}{ 17}{ 00}{668793}&     321& $-0.11\pm0.12$&$0.11\pm0.23$&$0.16\pm0.19$&     $+12.3\pm0.5$ & $-23.6\pm0.6$ & $26.6\pm 0.6$ & 0.10 & $3.32\pm0.08$ \\
                        &G2&\rahms{07}{ 42}{ 26}{562170}        &5.8&\decdms{-16}{ 17}{ 00}{668809}&       230     &\\
                        &G1+G2&\rahms{07}{ 42}{ 26}{562169} & 6.0& \decdms{-16}{ 17}{ 00}{668891} &214&\\
SAO\,135659&G1  &       \rahms{08}{ 13}{ 50}{953026}&   1.9&\decdms{-07}{ 38}{ 25}{622119} &      76&$-0.04\pm0.03$&$0.04\pm0.03$& $0.06\pm0.03$& $-10.5\pm1.2$ & $-72.6\pm 1.0$ & $73.3\pm 1.0$ & 0.14 & $3.67\pm  0.06$ \\
                        &G2&\rahms{08}{13}{ 50}{953037} &2.3&\decdms{-07}{ 38}{ 25}{622112}&       98&\\
                        &G1+G2& \rahms{08}{ 13}{ 50}{953031} & 1.7 &\decdms{-07}{ 38}{ 25}{622091} & 65 & \\
HD\,82159       &H1&\rahms{09}{ 30}{ 35}{542606}&       5.5     &\decdms{+10}{ 36}{ 05}{979934}&       151& $-0.04\pm0.03$&$0.02\pm0.03$&$0.04\pm0.03$&        $-1.24\pm0.2$ & $+0.2\pm 0.2$ & $1.3\pm 0.2$ &0.18& $ 0.05\pm 0.01$ \\
                        &H2     &\rahms{09}{ 30}{ 35}{542401}&  4.6     &\decdms{+10}{ 36}{ 05}{979351}&       145     &\\
                        &H1+H2&... & ... &... & ... & \\
HD\,82558       &H1&\rahms{09}{ 32}{ 25}{204609}&       5.9&\decdms{-11}{ 11}{ 03}{942747}&       185& $-0.07\pm0.29$&$0.03\pm0.45$&$0.07\pm0.31$&        $-0.6\pm0.2$ & $+0.2\pm 0.2$ & $0.7\pm 0.2$ &0.06& $0.010\pm  0.003$ \\
                        &H2&\rahms{09}{ 32}{ 25}{204348} &      4.9&\decdms{-11}{ 11}{ 03}{940469} &      183&    \\
                        &H1+H2&... & ... &... & ... &\\
HD\,199143&J1&\rahms{20}{ 55}{ 47}{759095}& 20&\decdms{-17}{ 06}{ 52}{304352}&  441& $0.47\pm0.15$&$0.48\pm0.15$&$0.67\pm0.15$&      $-2.6\pm0.3$ & $+3.6\pm 0.5$ & $4.4\pm 0.4$&0.27 & $0.20\pm  0.02$ \\
                        &J2     &\rahms{20}{ 55}{ 47}{759173}&  24&\decdms{     -17}{ 06}{ 52}{305210}& 608&\\
                        &J1+J2   &\rahms{20}{ 55}{ 47}{759121} & 15&\decdms{ -17}{ 06}{ 52}{304406} & 333&\\
GJ\,4199        &K1&\rahms{21}{ 31}{ 01}{925936}&       9.2 &\decdms{+23}{ 20}{ 04}{291167}&       144& $0.05\pm0.05$&$-0.06\pm0.05$&$0.08\pm0.05$&        $0.6\pm0.1$ & $+1.3\pm 0.2$ & $1.4\pm 0.2$ &0.19& $0.03\pm 0.01$ \\
                        &K2&\rahms{21}{ 31}{ 01}{925974}&       11  &\decdms{+23}{ 20}{ 04}{291129}&       190&    \\
                        &K1+K2&\rahms{21}{ 31}{ 01}{925953}& 6.7 &\decdms{+23}{ 20}{ 04}{291177}& 112 &\\
\arrayrulecolor{black}          \hline\hline
                        \label{tab:am}
                \end{tabular}
                \end{center}
\tablefoot{Columns are (from left to right) the target name, epoch ID as in Table~\ref{tab:Imr}, measured positions in the VLBA images and their statistical errors derived from JMFIT, and the estimated offset between the radio source detected with the VLA and the radio source detected with the VLBA ($\Delta\alpha^*_{\rm radio}$, $\Delta\delta_{\rm radio}$, and $\Delta\theta_{\rm radio}$). The offsets $\Delta\alpha^*$, $\Delta\delta$, and $\Delta\theta$ are the angular separations between the radio source (VLBA) and the {\it Gaia} DR3 predicted position at the observed epoch in right ascension, declination, and total, respectively.  The column $\sigma_{\rm sys}$ is the position systematic error expected from the phase difference of the target and its phase calibrator, given their angular separation. Finally, column
                $\Delta\lambda$, which is derived from $\Delta\theta$, is the linear separation in astronomical units at the corresponding distance (see Table~\ref{tab:tar}).}
\end{table*}

\section{Discussion}

\subsection{Comparison with VLA results}

The radio sources associated with our target stars are listed in Table~\ref{tab:tar}
and were first reported by \citet{launhardt2022}. These authors
measured the variability of the radio 
sources and pointed out that they are nonthermal emitter candidates,
motivating the follow-up VLBA observations presented here.  
We confirm the nonthermal nature of the radio sources detected with the VLBA, as they require brightness temperatures $>10^6\,$~K to be detected with interferometric baselines of $\sim$8\,000\,km \citep[see chapter 9 of ][]{thompson2017}. { These brightness temperatures are a strong indication of the nonthermal radio emission associated with young stars \citep{feigelson1999}. }

The angular resolution of VLA images presented by \citet[][]{launhardt2022} is  $\sim2''$, 
which is insufficient to resolve potential tight binaries.
As an example, in the case of EC~95 in Serpens,
VLA observations showed the presence of a single radio source \citep{eiroa2005}.
Follow-up VLBA observations showed that the radio source is a hierarchical
triple system, with three nonthermal variable radio emitters unresolved 
at the VLA angular resolution. The first VLBA observations of EC~95 showed a close
binary with an angular separation of about 15~mas \citep{dzib2010}, and later 
a third radio source was detected
at an angular distance of about 140 mas from the close binary \citep{ortiz2017b}.
Because of their variability, not all components were detected in all 
observed epochs. Thus, whether the radio sources detected in this work 
are identical to the radio sources reported by 
\citet{launhardt2022} needs to be analyzed. Indeed, the analysis by
\citet{launhardt2022} showed that, in some cases, the radio emission can have 
contributions from both components in known binary systems or be related 
to an unknown companion. 

\begin{figure}[ht!]
\centering
\includegraphics[width=0.45\textwidth, trim= 10 0 30 40,clip]{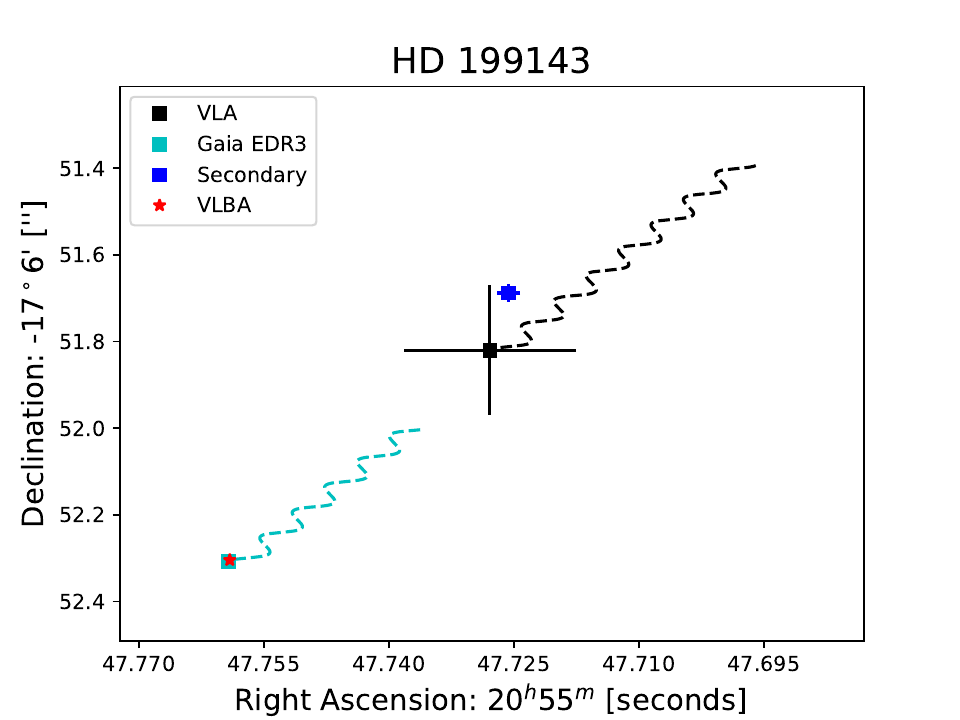}
\caption{Positions of the radio and optical sources related to HD 199143. 
The red star indicates the position of the radio source in the first 
detected epoch by the VLBA observations. The Cyan dashed line indicates the optical trajectory of the primary component of the HD 199143 stellar
system from 2016.0 
to the epoch of detection of the radio source with the VLBA. The optical position
of the star at the VLBA observed epoch is indicated with the cyan square. 
The blue square indicates the position of the secondary, HD\,199143\,B, whose absolute coordinates are unknown. Coordinates relative to the primary from the WDS catalog were used to estimate the absolute position.
{\it Gaia} DR3 and VLBA position errors are smaller than the symbol sizes.
The black line is the trajectory followed by the radio source detected
with the VLA \citep{launhardt2022}, assuming the derived astrometry at 
optical wavelengths from the VLA observed epoch
to the VLBA observed epoch. The black square indicates the 
extrapolated position of the VLA radio source at the epoch of the VLBA 
detection. The black cross size indicates the VLA positional error.   
}\label{fig:VVG}
\end{figure}

For our analysis, we compared the positions derived from the VLBA and the VLA.
First, we assumed that the radio sources share the trigonometric parallax
and proper motions of the associated stars reported by {\it Gaia} (see 
Table~\ref{tab:tar}) to extrapolate the positions reported from the VLA 
observations to the epoch of the corresponding first VLBA observation.
Then, we computed the offset between the position determined by the VLBA
observations and the extrapolated position from VLA observations 
($\Delta\alpha_{\rm radio}$, $\Delta\delta_{\rm radio}$ and $\Delta\theta_{\rm radio}$). The position offsets of these radio observations 
are  listed in {columns (7) to (9)} of Table~\ref{tab:am} and are shown graphically in Figures~\ref{fig:VVG} 
and \ref{fig:AmP}. 
Given the differences 
in angular resolution, the dominant error in the position offsets is 
the VLA position error, which is two orders of magnitude 
larger than those from the VLBA.

The offsets between the VLA and VLBA positions are less than three times the errors 
for nine out of the ten detected sources. In these cases, 
the VLA and VLBA radio sources presumably correspond to the same emitting source. 
In the case of HD\,199143, the separation is larger 
than three times the errors. This indicates that the radio source detected with 
the VLBA corresponds to a different object than the one reported from 
VLA observations. \citet{launhardt2022} argued that the radio source
detected in the VLA observations is positionally consistent with the expected 
position of the fainter companion, HD\,199143\,B (not in the {\it Gaia} DR3 catalog), 
located $0\rlap{.}''84$ northwest of 
the main stellar component \citep[see also Fig.~\ref{fig:VVG}]{hagelberg2020}. 
In the next section, we show that the radio source detected in the VLBA images 
corresponds to the primary stellar component in this system or to a very nearby companion.

\subsection{Comparison with optical astrometry}

\begin{table*}[t]
\small
\begin{center}
\renewcommand{\arraystretch}{1.0}
\caption{Previous companion information.}
\begin{tabular}{cccccccc }\hline\hline
            & Companion & Angular    & $\Delta\theta$& PMa$^{\rm a}$ & RUWE & SORS$^{\rm b}$ & SORS\\
Star ID     &  Name     & separation &  (mas) &  (SNR)   &  & &This work\\ 
\hline
V\,875\,Per & WDS\,02523+3617\,B$^{\rm c}$ & $5\rlap{.}''4$ & $0.6\pm0.2$&...&1.2&Main & Main\\
HD\,22213  & WDS\,03343--1204\,B & $1\rlap{.}''7$ & $1.1\pm0.3$ & ...  & 1.1&Main & Main \\  
HD\,285281 & WDS\,04005+1935\,B  & $0\rlap{.}''8$ & $0.9\pm0.5$  & ...  & 1.5& Main& Main \\
AI\,Lep    & WDS\,05403--1940\,B & $8\rlap{.}''4$ & $0.5\pm0.2$  & ...  & 1.0 & Main & Main\\
\rowcolor{lightgray}
HD\,62237  & ...              & ...            & $26.6\pm0.6$  & ...  &7.3  & Main& {New companion}\\
\rowcolor{lightgray}
SAO\,135659& WDS\,08138--0738\,B & $0\rlap{.}''1$& $73.3\pm1.0$   & ...& 13.8 &Main/Companion& Companion\\
HD\,82159  & WDS\,09306+1036\,B  & $13\rlap{.}''8$& $1.3\pm0.3$ & 1.74  & 1.6 & Main& Main\\
HD\,82558  & ...              & ...            & $0.7\pm0.2$  & 0.79 & 1.1  & Main&  Main \\
\rowcolor{lightgray}
HD\,199143 & HD\,199143\,B      & $0\rlap{.}''84$ & $4.4\pm0.5$  & 56.68 & 1.0&Companion& New Companion \\
GJ\,4199    & WDS\,21310+2320\,B & $9\rlap{.}''2$ & $1.4\pm0.3$ & 0.99  & 1.0 &Main& Main\\
\arrayrulecolor{black}\hline\hline
\label{tab:bin}
\end{tabular}
\end{center}
\tablefoot{Angular separations of previously known companions
are from the Washington Visual Double Star Catalog
\citep{mason2001,mason2022}, except for HD 199143, which is from \citep{hagelberg2020}. The quoted error value in $\Delta\theta$ was estimated by adding in quadrature the statistic and systematic errors listed in Table~\ref{tab:am}.\\
Sources discussed further in the text are highlighted.\\
\tablefoottext{${\rm a}$}{As reported by \citet{kervella2019} or \citet{kervella2022}.}
\tablefoottext{${\rm b}$}{Suggested origin of the radio source (SORS) as reported by \citet{launhardt2022}.} 
\tablefoottext{${\rm c}$}{No physical companion \citep{mason2022}.}}
\end{table*}

To examine the relationship between the radio sources detected with the VLBA 
and the targeted stars, we have used the astrometric results from optical 
wavelengths (i.e., Gaia)
to determine the stellar positions at the epoch of the VLBA
observations. Then, we determined the angular offsets
between the radio source and the optical counterpart 
($\Delta={\rm Pos}_{\rm radio}-{\rm Pos}_{\rm Opt.}$).
The initial parameter uncertainties (optical: position, proper motion, and parallax; 
radio: position) were propagated to estimate the angular separation error 
by using a Monte Carlo approach. In columns { 10, 11, and 12} 
of Table~\ref{tab:am}, we list the angular separation 
in right ascension, $\Delta\alpha*=\Delta\alpha\cdot\cos{({\rm Dec.})}$ and in
declination, ($\Delta\delta$), and the total angular separation
($\Delta\theta$) between the VLBA and optical sources. Expected systematic errors in the source position 
caused by residual phase calibration errors are listed in column 11 and were added quadratically to the errors provided by JMFIT. 

Seven of the ten radio sources are consistent with the
expected position from {\it Gaia} results within five times the errors. 
We argue that, in these cases, the radio emission is tracing the stellar corona of the stars seen at optical wavelengths. Stellar coronae of young stars could be of sizes up to 10 R$_*$ \citep[see the review by][]{gudel2002}. For a conservative value for the stellar radii, we assumed 1 R$_*$= 1.15 R$_\odot$, as the observed stars are of spectral types F8 or later and about to enter the main sequence. At the distances of our targets, this implies angular sizes of  less than $2.9$ mas.
On the other hand, the measured positions of the other three radio sources (HD\,62237, SAO\,135659, and HD\,199143) are not consistent with
the results from {\it Gaia}, and the offset cannot be attributed to the stellar coronae structure, as they would require structures greater than $40$~R$_*$.

To look for clues on the origin of the radio-optical position discrepancies in these three sources, we first turned to the {\it Gaia} results. The {\it Gaia} catalog includes a measure of the quality of the astrometric fits called the 
renormalized unit weight error (RUWE; listed Table~\ref{tab:bin} for our targets). The RUWE index is expected to be close to 1.0 for stars whose 
motion is well described by a uniform linear motion, while a large RUWE value ($>1.4$) indicates a significant deviation from a linear uniform motion and is suggestive of the presence of an unseen companion. Interestingly, two of the three sources where we found discrepancies between the VLBA and optical positions (HD\,62237 and SAO\,135659) have very large RUWE values (7.3 and 13.8, respectively). A second relevant piece of information is provided by the proper motion anomaly (PMa) introduced by \citet{kervella2019} and \citet{kervella2022} and listed in the fifth column of Table \ref{tab:bin}. The PMa quantifies discrepancies between the proper motions measured by the Hipparcos and {\it Gaia} astrometric missions. A large PMa indicates significantly different proper motions in the Hipparcos and {\it Gaia} catalogs and suggests the presence of a low-mass companion with an orbital period larger than the {\it Gaia} observation window (668 days for {\it Gaia} DR3). The third source where we found a discrepancy between the VLBA and optical positions (HD\,199143) has a very large PMa (56.7). 

In summary, the three sources where the VLBA and optical positions do not coincide have astrometric evidence of multiplicity. In contrast, the sources where the VLBA and optical positions do coincide have RUWE values close to unity and either small PMa values or no PMa at all. 

Given the evidence presented above that the sources with discrepant VLBA and optical positions are multiple, our next step was to look for known companions. 
We searched the Washington Visual Double Star Catalog \citep[WDS; ][]{mason2001,mason2022} for companions to our target stars.\footnote{For completeness, we searched the WDS catalog for companions to all of our targets, not just for HD 62237, SAO 135659, and HD 199143.} We list the stellar companions and their angular separation in Table~\ref{tab:bin}. In cases where there is more than one known companion, we include only the companion with the smallest angular separation to the targeted star. For HD 62237 and HD 199143, the angular separation of the visual binaries is much larger than the VLBA-optical astrometric discrepancy $\Delta\theta$. In the case of SAO\,135659, however, the values are comparable. We come back to this point momentarily. We also investigated if our targets had been reported as spectroscopic binaries. Four of our target sources (HD\,22213, SAO\,135659, GJ\,4199, and HD\,199143) were included in the spectroscopic analysis by \citet{zuniga2021}. Only HD\,22213 was confirmed as a spectroscopic binary. In particular, none of the stars with discrepant optical and VLBA positions are a known spectroscopic binary.

Armed with the information gathered for each system, we are now in a position to discuss the three targets where the VLBA and optical positions do not coincide. {Particularly, we discuss if the origin of the position discrepancy can be explained by known companions.}

\noindent
Firstly, SAO\,135659 is known to be a binary star with stellar components of similar masses (0.77 and 0.64 M$_\odot$ for the primary and secondary, respectively). The angular separation of the system is $\sim0\rlap{.}''1$ almost in the north-south direction at the epoch of the observations reported by \citet{elliott2015}. At a distance of 50 pc and given the total mass, we can estimate the period of the system to be roughly 9 years. Coincidentally, this is similar to the separation between the observations realized by \citet{elliott2015} in May 2012 and our VLBA observations in April 2021. Consequently, we expect the relative offset between the two stars to also be on order of $\sim0\rlap{.}''1$ in the north-south direction at the time of our VLBA observations. In our observations, the position of the VLBA source was found to be located about $\sim0\rlap{.}''07$ north of the expected {\it Gaia} position (Fig.~\ref{fig:AmP}). The similarity between the relative position of the optical and VLBA sources and the expected relative positions of the two stars in SAO 135659 strongly suggests that the VLBA source is associated with the secondary (WDS 08138-0738). The small difference in separation ($\sim0\rlap{.}''1$ vs.\ $\sim0\rlap{.}''07$) could easily be caused by the somewhat different orbit phase seen during our observations compared to those of \citet{elliott2015} and by the fact that {\it Gaia} presumably traces the position of the photocenter of the system rather than the exact position of the primary.\\

\noindent
The source HD\,199143 exhibits a large PMa. The detailed analysis from \citet{launhardt2022} shows that this can be explained by the known companion HD\,199143\,B located at $\sim0\rlap{.}''84$ \citep[see Fig. 12 from][]{launhardt2022}. The VLA source reported by \citet{launhardt2022} coincides with that companion (see also Fig.\ref{fig:VVG}). The VLBA source we detect here, however, is located very near the primary of the system (Fig.\ref{fig:VVG}). The small but significant offset between the VLBA position and the expected position of the primary (4.4 $\pm$ 0.5 mas) could be due to the presence of another hitherto undetected companion located very close to the primary. Given the small offset, however, further observations will be needed. To finish on this source, we note that we searched for a second VLBA radio source in the expected position of the companion but found none. The flux of the companion star was below 110 $\mu$Jy~ beam$^{-1}$(6.5 times the noise level in our image) during the VLBA observations. {The radio source detection by \citet{launhardt2022} and the one reported here is interesting but unsurprising, and it points to the highly variable nature of the radiation mechanisms in magnetically active stars \citep[e.g.,][]{gudel2002}. }

Regarding HD\,62237, it has no known companion, but the large {\it Gaia} RUWE value strongly suggests that such a companion exists. We argue that the VLBA source, offset by about 27 mas from the expected position of the primary, traces that companion. We note that, given the modest resolution of the VLA observations reported by \citet{launhardt2022}, the VLA source could coincide with either the primary or the secondary putatively traced by the VLBA (Fig.\ref{fig:VVG}).
{This companion should be less luminous than the primary star, as a clear offset between the Gaia and VLBA positions indicates that the system 
photocenter is coincident or close to the primary star. However, the mass of the companion must be significant to cause the linear fitting by Gaia DR3 to not properly describe the motion of the system (RUWE value is 7.3). The spectral type of HD\,62237 is G5V, indicating a star with a mass of 0.93 M$_\odot$ \citep{zombeck1990}. Assuming that the companion has a mass between 0 and 0.5 M$_\odot$, the mass of the system ranges from 0.93 to 1.43 M$_\odot$. 
We also assume that the optical-radio offset of $26.6\pm0.6$ mas ($3.32\pm0.08$\,au at a distance of 124.76 pc) is the semi-major axis of the orbit, and we roughly estimate an orbital period of the system in the range from 5.1 to 6.3 years.}


\section{Conclusions}

We have presented a series of VLBA observations of a sample of 31 nearby young stars previously reported 
as radio sources, obtaining ten detections (for a detection rate on the order of 30\%). The radio sources are compact  
at our angular resolution of a few milli-arcseconds and are highly variable; both properties confirm they are nonthermal radio emitters. All detected radio sources are associated with stars with good astrometry from {\it Gaia} DR3.
Using these results, we extrapolated the position
of the optical stars to the epoch of the radio 
observations to compare the radio and optical positions.
For seven of the ten detected
radio sources, the optical and radio positions are consistent. Three radio sources, however, have 
position discrepancies above eight times the errors. All three of these sources show independent evidence of multiplicity, and we attribute the position discrepancies to the presence of companions.
 
Follow-up observations of the detected radio sources 
will be of high interest for comparison with {\it Gaia} astrometry. Such observations could provide some clues on the {\it Gaia} 
parallax zero point and on the relative orientation of the {\it Gaia} and ICRF reference frames {by increasing the current numbers of stars with Gaia- and VLBI-derived astrometry}.

\begin{acknowledgements}
We thank the anonymous referee for his/her careful review and insightful comments.
We thank Eduardo Ros for reading the paper and providing suggestions that improve the readability of the paper. 
S.A.D. acknowledge the M2FINDERS project from the European Research
Council (ERC) under the European Union's Horizon 2020 research and innovation programme
(grant No 101018682). L.L. acknowledges the support of UNAM-DGAPA PAPIIT grants IN108324 and  IN112820 and 
CONACYT-CF grant 263356. J.O.-T. acknowledges a stipend from CONAHCYT, M\'exico.
The National Radio Astronomy
Observatory is a facility of the National Science Foundation
operated under cooperative agreement by Associated 
Universities, Inc.

This work has made use of data from the European Space 
Agency (ESA) mission {\it Gaia}
(\url{https://www.cosmos.esa.int/gaia}), processed by the 
{\it Gaia} Data Processing and Analysis Consortium (DPAC,
\url{https://www.cosmos.esa.int/web/gaia/dpac/consortium}). 
Funding for the DPAC has been provided by national 
institutions, in particular the institutions
participating in the {\it Gaia} Multilateral Agreement.
\end{acknowledgements}

\bibliographystyle{aa}
\bibliography{references}

\begin{appendix}

\section{Images of detected radio sources}\label{app:1}

We detected ten of our 31 targets. The obtained images from our VLBA 
observations are presented in this appendix. The plots in Figure~\ref{fig:NF2}
are the intensity maps of nine detected radio sources; the remaining source
is presented in the main text. The figures include the expected positions
of the stars at the epochs of the VLBA observations as calculated from {\it Gaia} DR3 \citep{gaiadr32021,gaiadr3_2023}. 

To produce the images combining the data from both epochs, we corrected the position of the second epoch, assuming that the optical astrometric parameters could be applied to the radio observations. 
In AIPS, this was done with the task {\it CLCOR} using the operation code {\it ANTC}. The position correction for each target was performed by specifying the offsets in right ascension and declination in arcsecond units ({\it clcorprm(5,6) }=$\alpha_{\rm offset}$, $\delta_{\rm offset}$). The values of these offsets are listed in Table~\ref{tab:off}.

We assumed the upper limits for the nondetected sources to be 6.5 times the noise levels
of images where the two observed epochs of each source were combined. These noise level
values are also listed in Table~\ref{tab:off}. 

\begin{table}[h!]
\small
\begin{center}
\renewcommand{\arraystretch}{1.0}
\caption{Offsets applied to combine data of the two observed epochs and the noise level of the obtained image.}
\begin{tabular}{cccc}\hline\hline
            &  \multicolumn{2}{c}{Offsets (mas)}& Noise\\
Star ID     &  $\alpha$    & $\delta$ &$\mu$Jy bm$^{-1}$\\ 
\hline
BD17232 & $-0.45$ & $+0.13$ & 25 \\ 
HIP\,12545 &$-0.30$ & $+0.22$ & 13\\   
HIP\,12635 &$-0.41$ & $+0.89$ & 14 \\
\rowcolor{lightgray}
V\,875\,Per & $-0.32$ & $+0.27$& 23\\ 
TYC\,3301  & $-0.37$ & $+0.79$& 15 \\
\rowcolor{lightgray}
HD\,22213 & $-0.55$ & $-0.23$ & 17 \\ 
HD\,23524 & $-0.44$ & $+0.76$ & 19 \\ 
HD\,24681 & $-0.21$ & $+0.43$ & 12 \\ 
\rowcolor{lightgray}
HD\,285281 & $+0.02$ & $+0.07$ &13 \\  
HD\,284135 & $+0.05$ & $+0.11$ & 14 \\ 
HD\,31281 & $-0.00$ & $+0.03$ & 21 \\ 
BD\,08995 & $-0.08$ & $-0.05$ & 16 \\ 
HD\,286264 & $-0.06$ & $+0.12$ & 16 \\ 
HD\,293857 & $-0.05$ & $+0.06$ & 15 \\ 
TYC\,7136611 & $+0.01$ & $+0.25$ & 10 \\ 
TYC\,59251547 & $-0.04$ & $-0.03$ & 12 \\ 
\rowcolor{lightgray}
AI\,Lep & $-0.04$ & $-0.13$ & 12 \\ 
\rowcolor{lightgray}
HD\,62237 & $+0.05$ & $-0.10$ & 15 \\ 
\rowcolor{lightgray}
SAO\,135659 & $+0.26$ & $-0.15$ & 16 \\ 
WDS\,09035+3750B & $+0.42$ & $+4.74$ & 19 \\ 
\rowcolor{lightgray}
HD\,82159 & $+3.42$ & $+0.14$ & 38 \\ 
\rowcolor{lightgray}
HD\,82558 & $+4.01$ & $-3.24$ & 22 \\ 
GJ\,2079 & $+2.75$ & $+4.44$ & 16 \\ 
HD\,135363 & $+1.02$ & $-1.03$ & 20 \\ 
UCAC\,4832 & $+1.02$ & $-1.03$ & 14 \\ 
\rowcolor{lightgray}
HD\,199143 & $-0.37$ & $+0.51$ & 17 \\ 
HD\,358623 & $-0.38$ & $+0.54$ & 16 \\ 
SAO\,50350 & $-0.35$ & $-0.93$ & 14 \\ 
\rowcolor{lightgray}
GJ\,4199 & $-0.28$ & $-0.00$ & 14 \\ 
SAO\,51891 & $-0.43$ & $-0.45$ & 18 \\ 
SAO\,108142 & $-0.29$ & $+0.05$ & 13 \\ 
\arrayrulecolor{black}\hline\hline
\label{tab:off}
\end{tabular}
\end{center}
\tablefoot{Offsets are as expected assuming the parallax and proper motions
obtained from the optical results. Detected sources are highlighted with gray-shaded rows. The noise value was measured in the 
image of two combined epochs, that is, after applying the correction of the 
offsets. 
}
\end{table}

\begin{figure*}[!h]
\centering
\includegraphics[width=0.98\textwidth, trim= 10 20 0 0,clip]{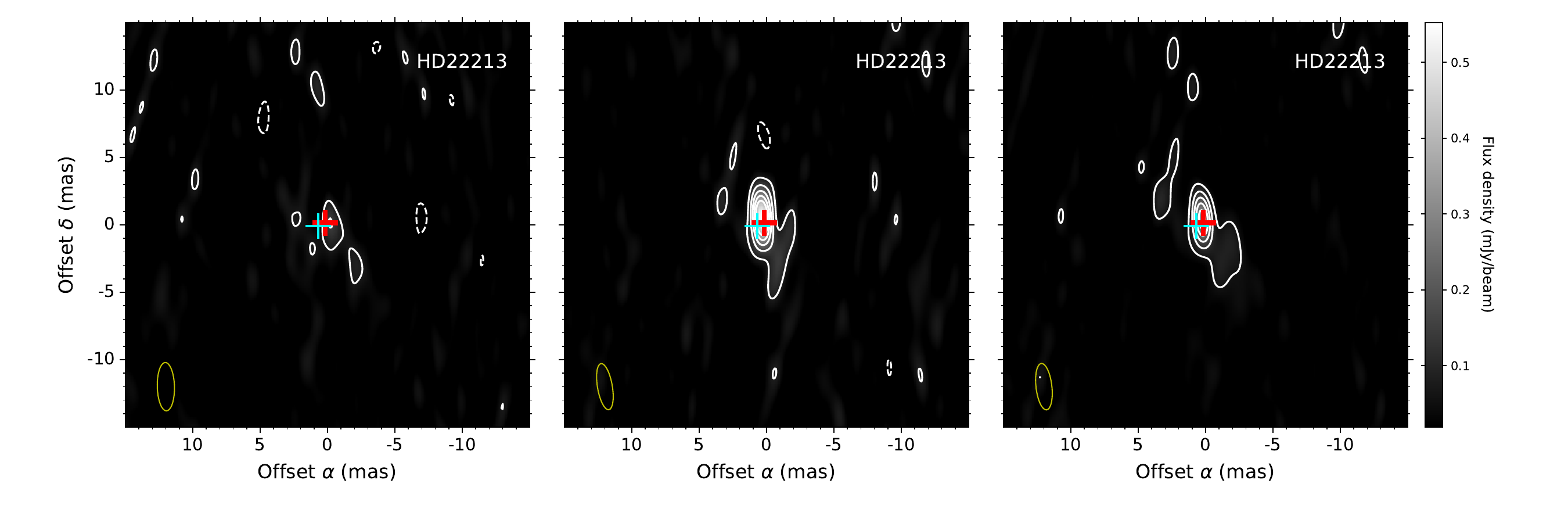}
\includegraphics[width=0.98\textwidth, trim= 10 20 0 0,clip]{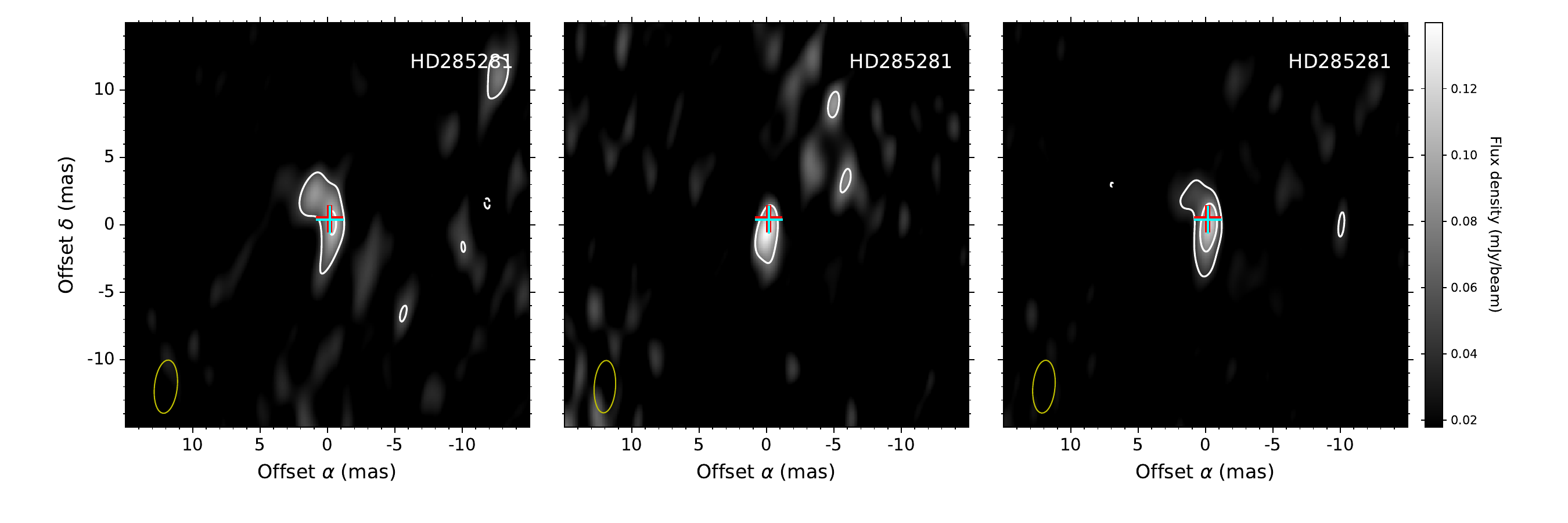}
\includegraphics[width=0.98\textwidth, trim= 10 20 0 0,clip]{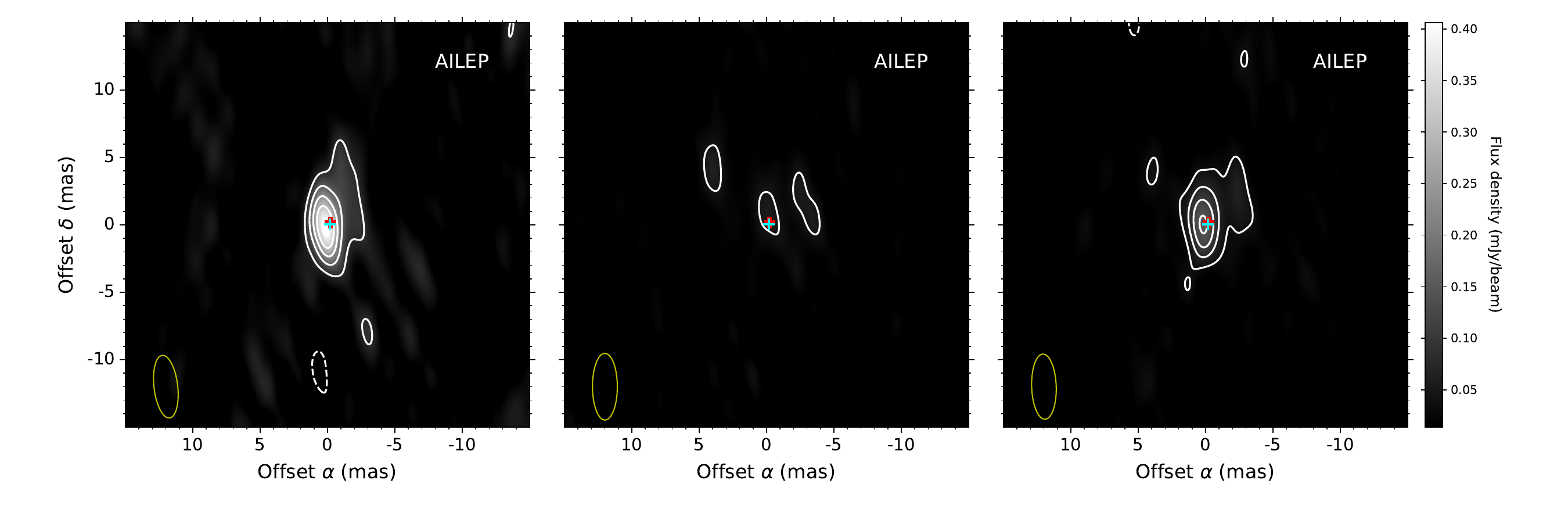}
\includegraphics[width=0.97\textwidth, trim= 10 20 0 0,clip]{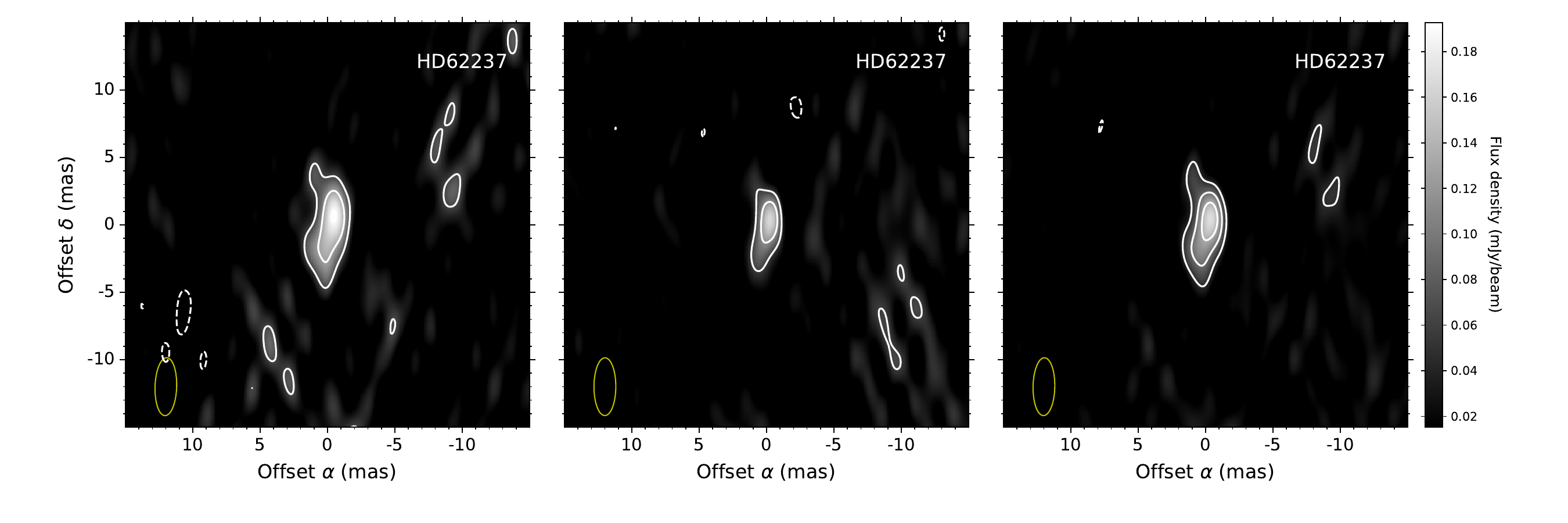}
\caption{VLBA images of detected radio sources. The images are centered in the position of the radio source as detected in the first epoch (see Table~\ref{tab:am}). The name of the stellar source related to the radio sources is indicated in the top-left corner. The images, from
left to right, correspond to the epochs 1, 2, and the combination of both epochs. Contour levels are -3, 3, 6, 9, 12, and 15 times the noise level of the image as listed in Table~\ref{tab:Imr}. The 
yellow open ellipse in the bottom-left corner represents the size of the synthesized beam of the image as listed in Table~\ref{tab:Imr}. The predicted optical position in epochs 1 and 2 are shown as red and cyan crosses, respectively. For HD\,62237 and SAO\,135659, these positions fall outside the region shown.}\label{fig:NF2}
\end{figure*}

\setcounter{figure}{0}
\begin{figure*}[!h]
\centering

\includegraphics[width=0.99\textwidth, trim= 10 20 0 0,clip]{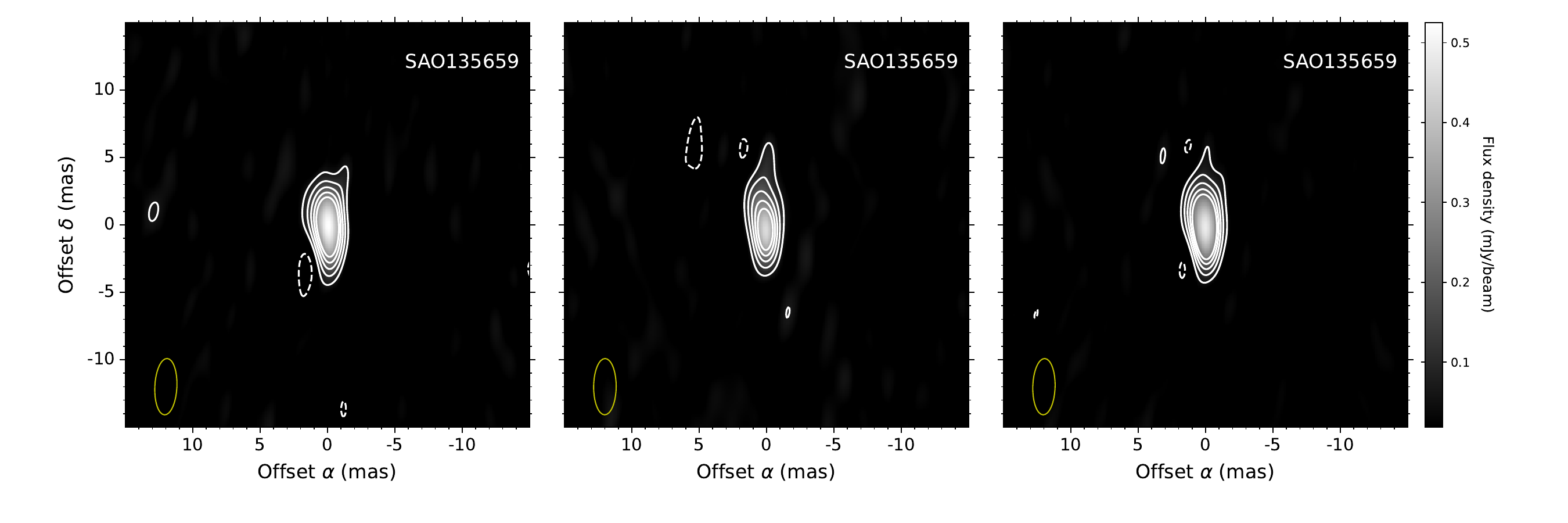} 
\includegraphics[width=0.98\textwidth, trim= 10 20 0 0,clip]{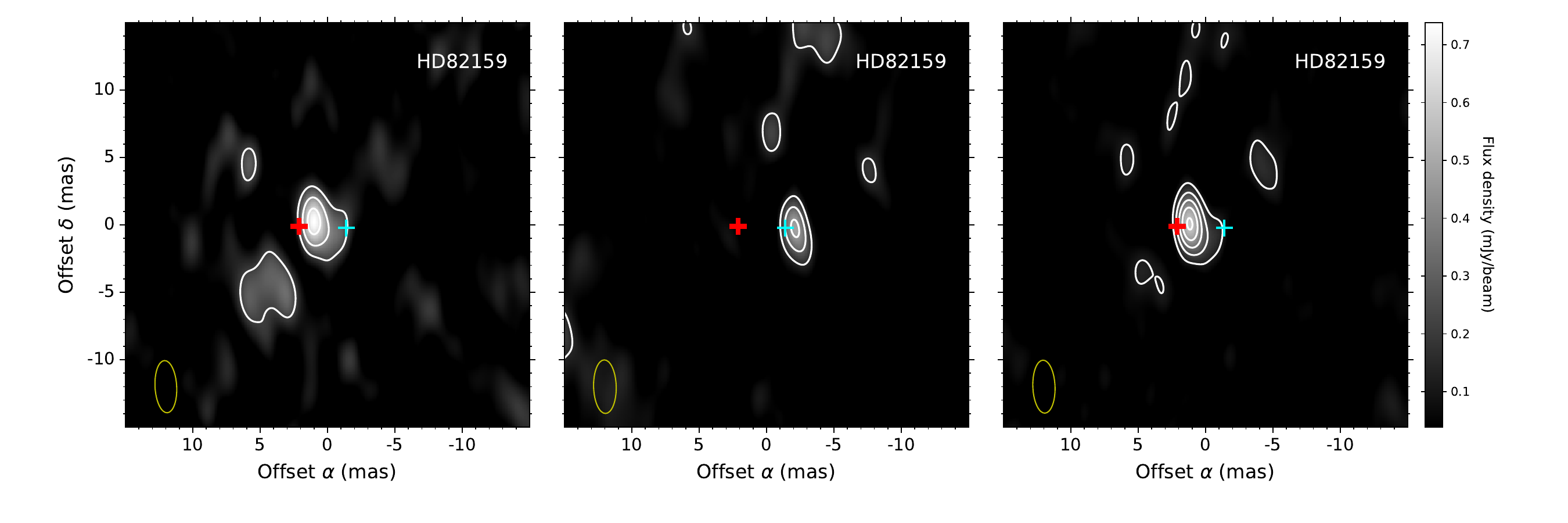} 
\includegraphics[width=0.98\textwidth, trim= 10 20 0 0,clip]{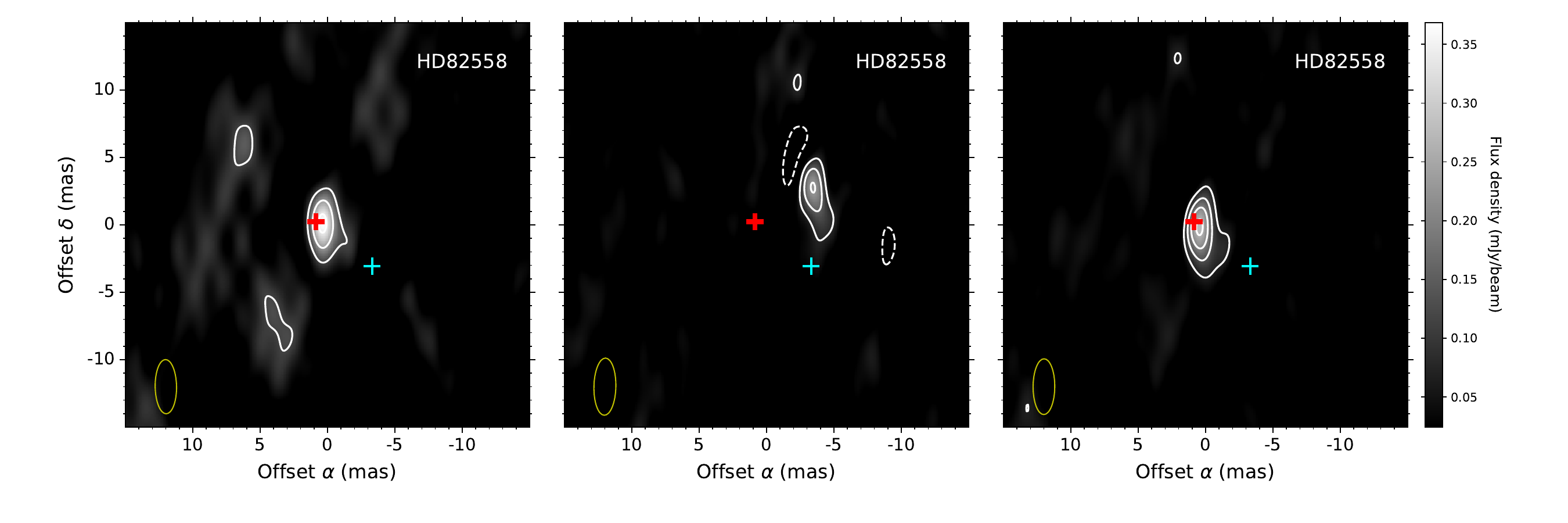}
\includegraphics[width=0.98\textwidth, trim= 10 20 0 0,clip]{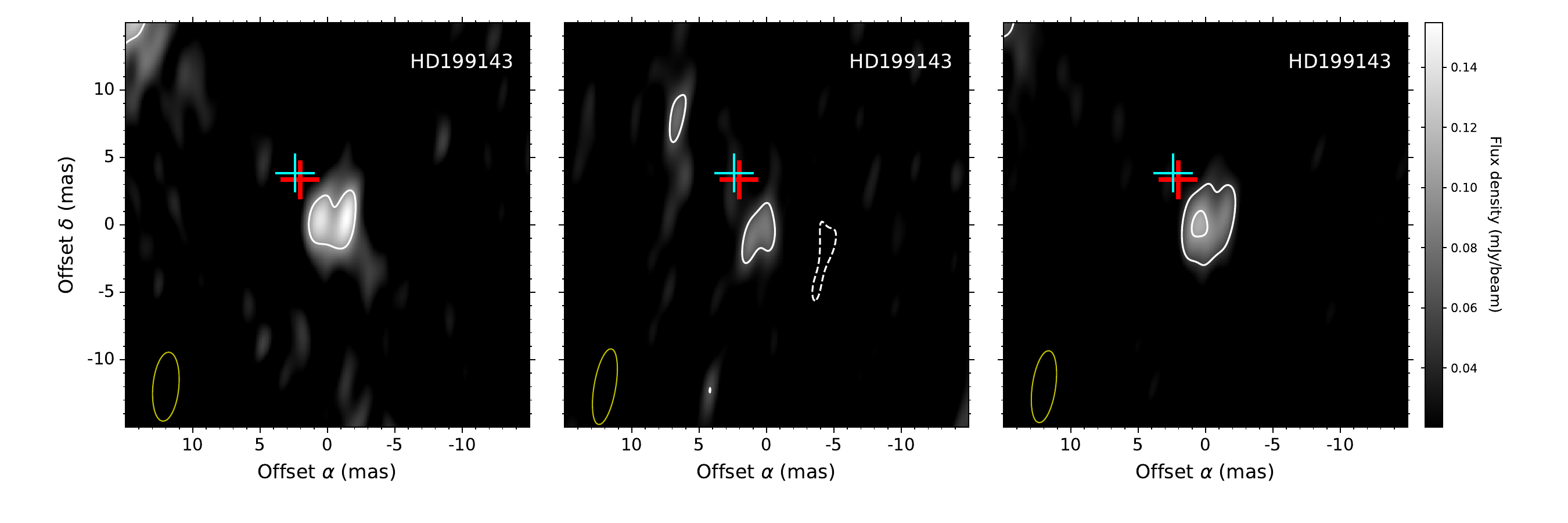}
\caption{Continued.}
\end{figure*}

\setcounter{figure}{0}
\begin{figure*}[!h]
\centering

\includegraphics[width=0.99\textwidth, trim= 10 20 0 0,clip]{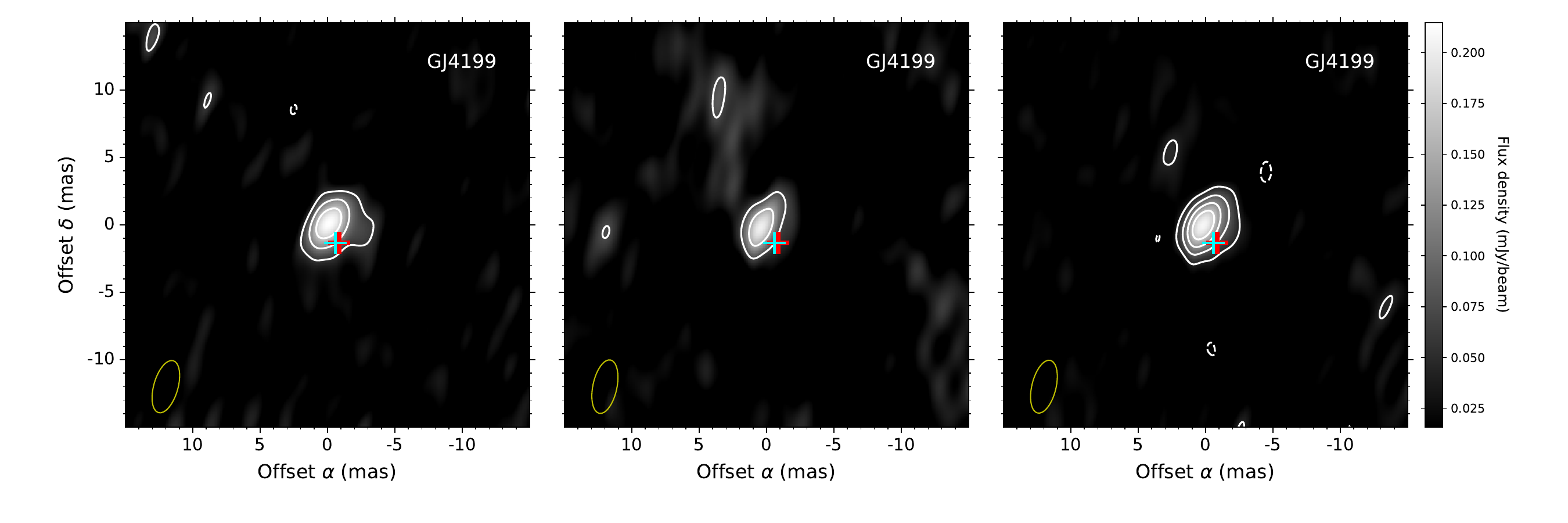}
\caption{Continued.}
\end{figure*}

\newpage

\section{Figures comparing VLBA, VLA, and \textit{Gaia} astrometric results}

\begin{figure*}[!h]
\begin{tabular}{ccc}
\centering
\includegraphics[width=0.32\textwidth, trim= 0 0 10 0,clip]{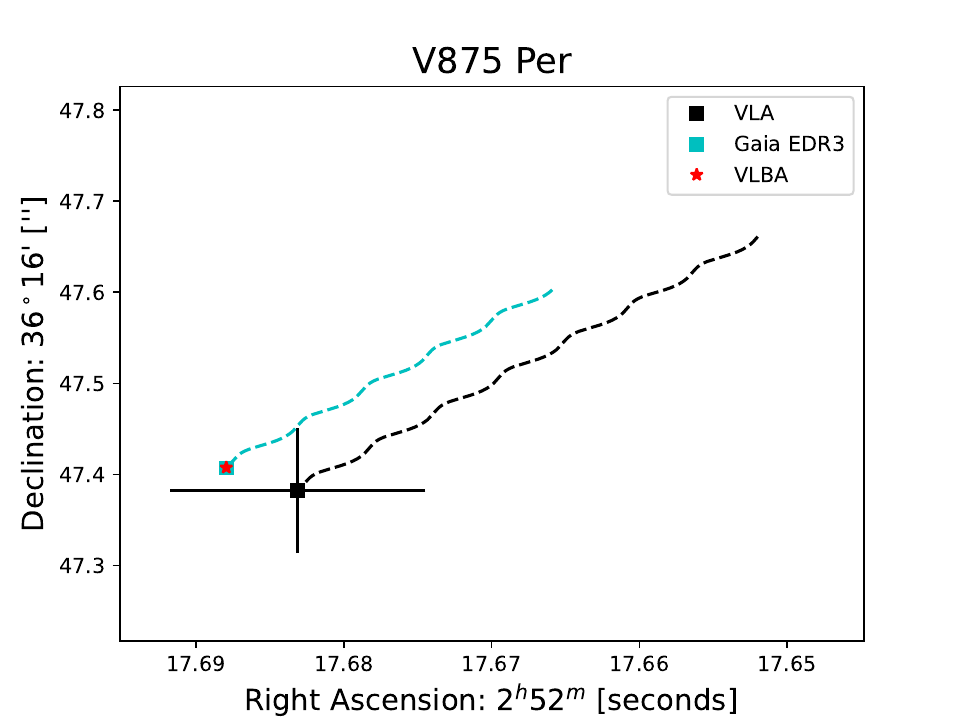} &
\includegraphics[width=0.32\textwidth, trim= 0 0 10 0,clip]{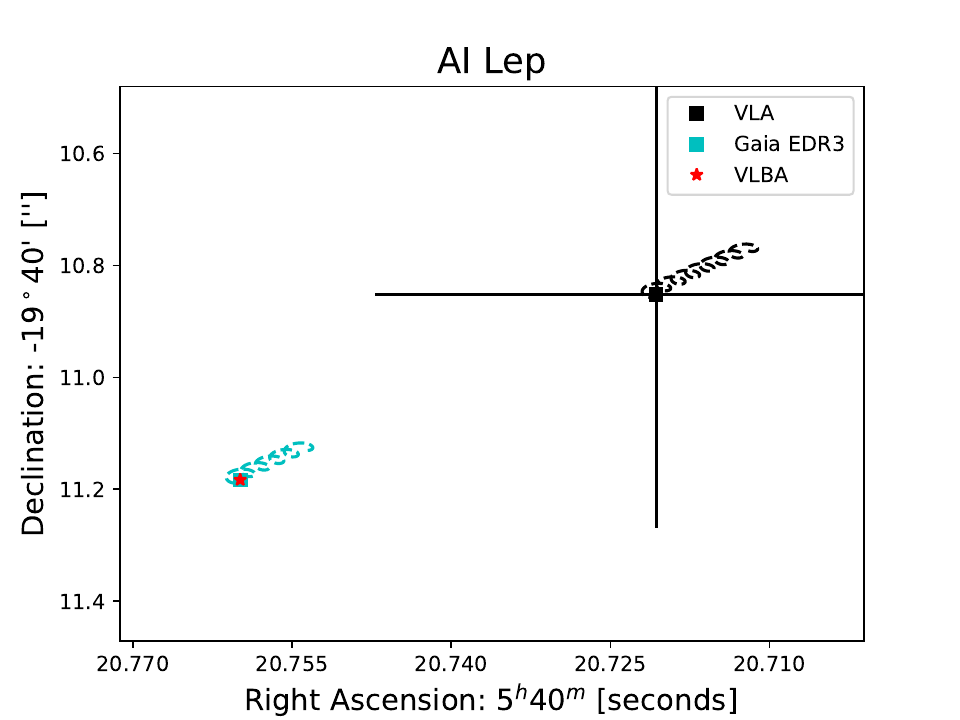} & \includegraphics[width=0.32\textwidth, trim= 0 0 10 0,clip]{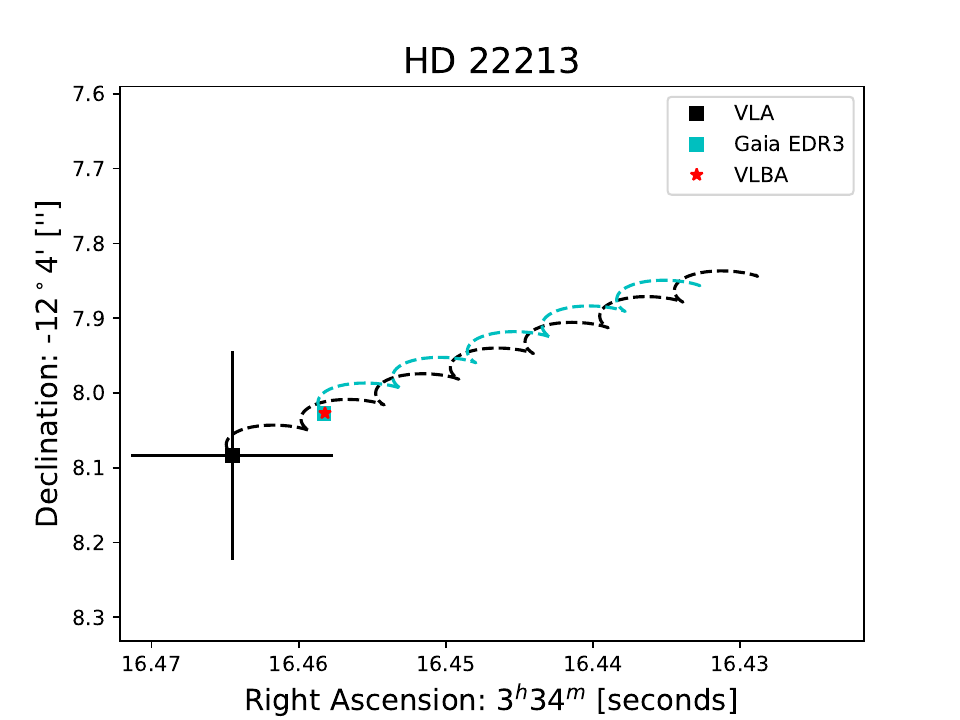}\\
\includegraphics[width=0.32\textwidth, trim= 0 0 10 0,clip]{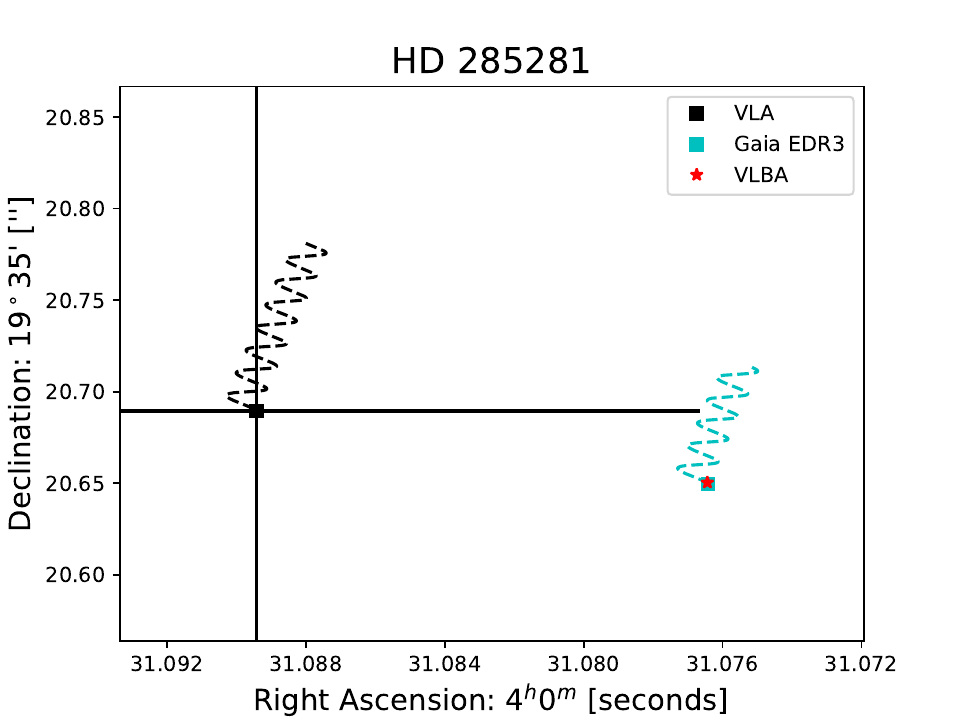} &
\includegraphics[width=0.32\textwidth, trim= 0 0 10 0,clip]{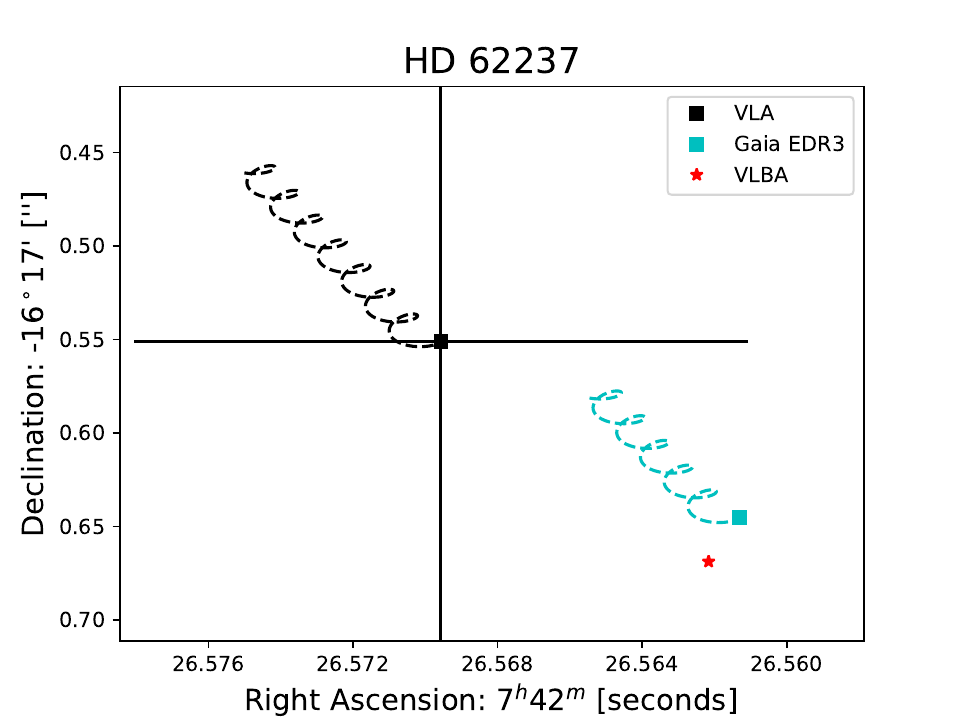} &
\includegraphics[width=0.32\textwidth, trim= 0 0 10 0,clip]{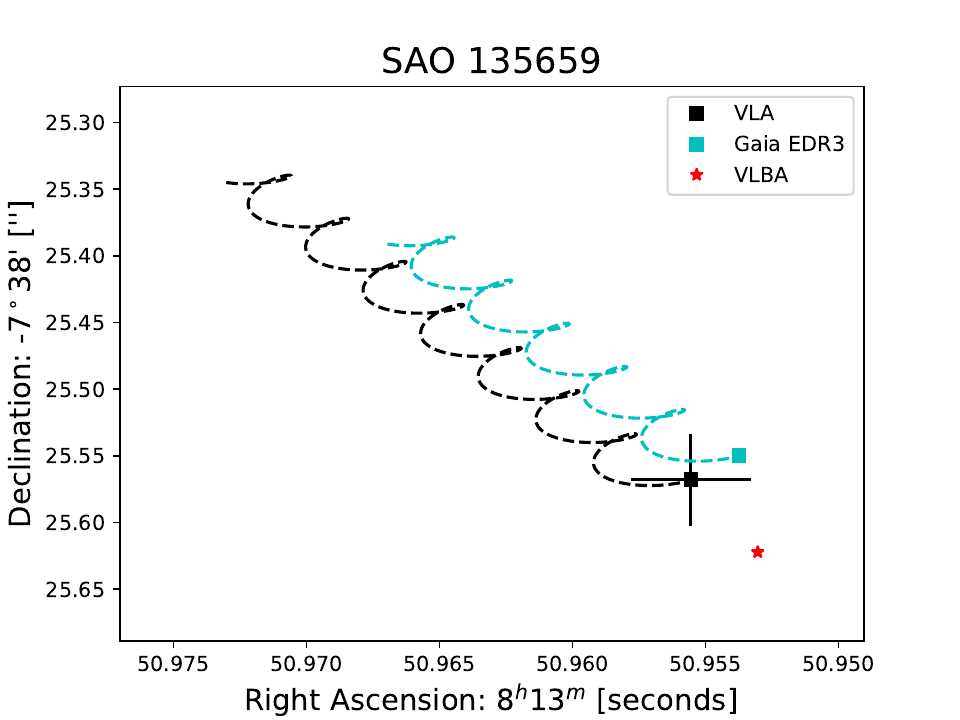} \\
\includegraphics[width=0.32\textwidth, trim= 0 0 10 0,clip]{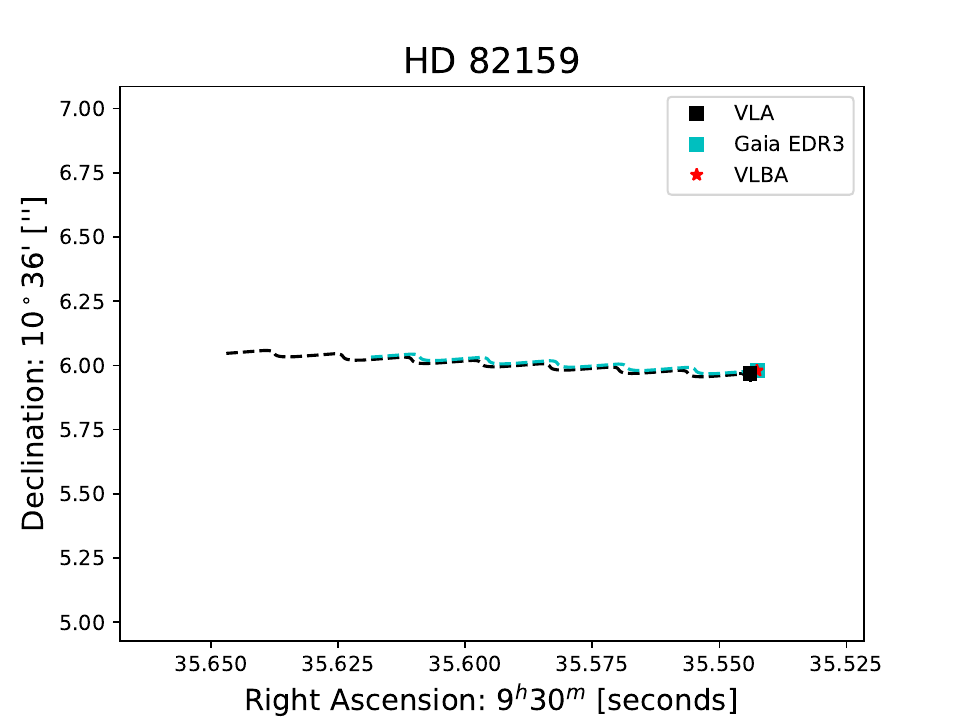}&
\includegraphics[width=0.32\textwidth, trim= 0 0 10 0,clip]{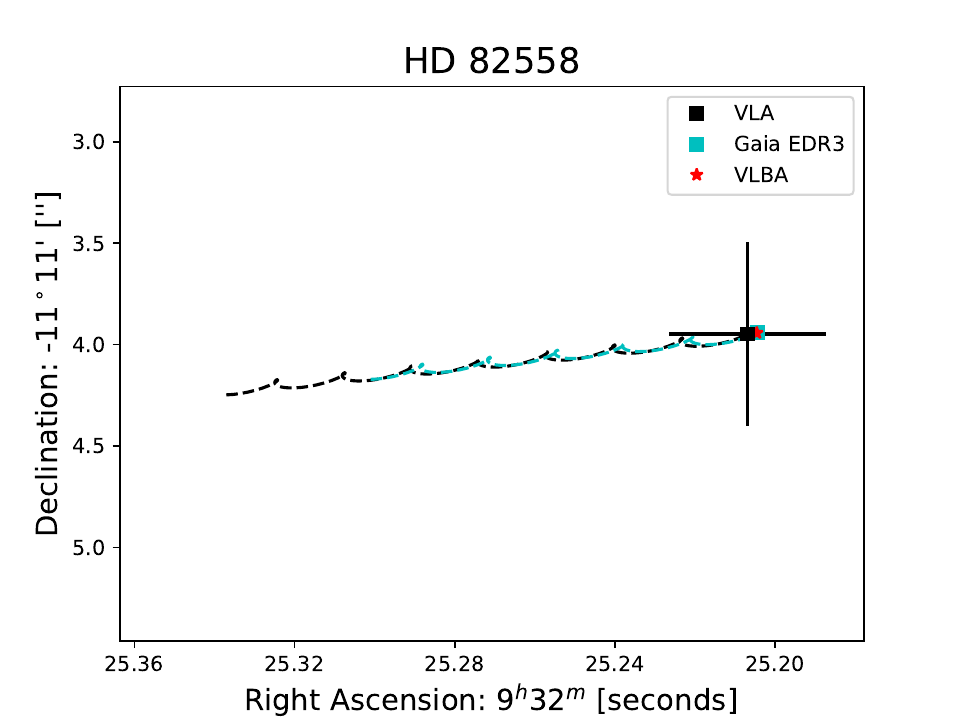} &
\includegraphics[width=0.32\textwidth, trim= 0 0 10 0,clip]{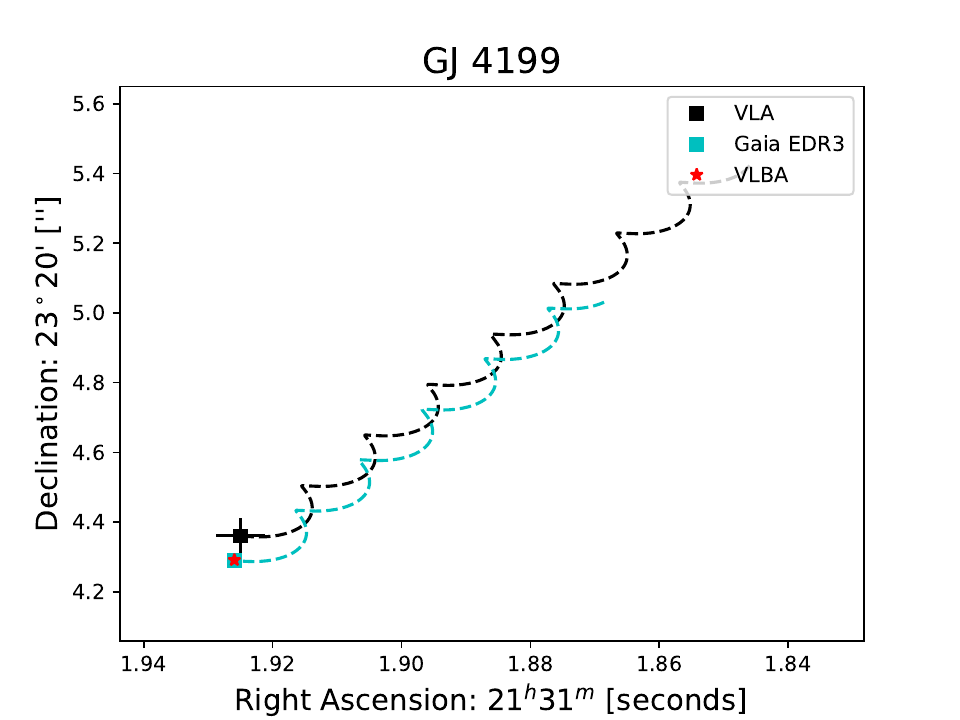}
\end{tabular}
\caption{Positions of radio and optical sources related to our target sources.
The name of source is indicated at the top of each plot.
Red stars indicate the position of the radio source in the first 
detected epoch by the VLBA observations. 
The black line is the trajectory followed by the radio source detected
with the VLA \citep{launhardt2022}, assuming the derived astrometry at 
optical wavelengths (see Table~\ref{tab:am}) from the VLA observed epoch
to the VLBA observed epoch. The black square indicates the 
extrapolated position of the VLA radio source at the epoch of the VLBA 
detection. The black cross size indicates the VLA positional error.   
The cyan line indicates the trajectory of the optical results from 2016.0 
to the epoch of detection of the radio source with the VLBA; the position
of the optical source at this epoch is indicated with the cyan square. {\it Gaia} DR3 and VLBA 
position errors are smaller than the symbol sizes. Companion relative positions are shown 
as blue squares.}\label{fig:AmP}
\end{figure*}

Figure~\ref{fig:AmP} shows the measured VLBA position of detected radio sources.
In these plots, for comparison, we also show the position of the VLA 
radio sources detected by \citet{launhardt2022}, where we have extrapolated 
its position to the epoch of the VLBA observation using the results of optical
astrometry. The expected optical position at the epoch of VLBA observations
is also shown. Finally, in two cases, SAO 135659 and HD199143, the relative 
position to the primary of a companion was obtained from the Washington Visual 
Double Star catalog \citep{mason2001,mason2022} and are also shown in these plots.

\end{appendix}

\end{document}